\documentclass[twocolumn, twocolappendix]{aastex701}
\usepackage{threeparttable}
\usepackage{multirow}
\usepackage{hyperref}
\usepackage{xcolor}
\usepackage{soul}

\begin{document}

\title{Spin-Orbit Geometry of AU Mic b and c from Back-to-Back Transits Observed\\
Contemporaneously with Magellan PFS, LCOGT, and CHEOPS\footnote{This paper includes data gathered with the 6.5 meter Magellan Telescopes located at Las Campanas Observatory, Chile, and CHEOPS data observed as part of Guaranteed Time Observations (GTO) programme CH\_PR140071 (PI: Szabo).}}

\correspondingauthor{Zitao Lin, Zhen Guo, Sharon X. Wang}
\email{lzt22@mails.tsinghua.edu.cn, zhen.guo@uv.cl, sharonw@mail.tsinghua.edu.cn}

\author[0000-0001-5695-8734]{Zitao Lin}
\affiliation{Department of Astronomy, Tsinghua University, Beijing 100084, China}
\affiliation{Chinese Academy of Sciences South America Center for Astronomy (CASSACA), National Astronomical Observatories, CAS, Beijing 100101, China}
\email{lzt22@mails.tsinghua.edu.cn}

\author[0000-0002-0606-7930]{Gyula M. Szabó}
\affiliation{ELTE Gothard Astrophysical Observatory, H-9700 Szombathely, Szent Imre h. u. 112, Hungary}
\email{szgy@konkoly.hu}

\author[0009-0001-0389-8907]{Krzysztof Sz. Zieliński}
\affiliation{Obserwatorium Astronomiczne Niedźwiady, Szubin, Poland}
\affiliation{Boyce Research Initiatives and Education Foundation, San Diego, CA, USA}
\email{krzysztof.sz.zielinski@gmail.com}

\author[0000-0003-0292-4832]{Zhen Guo}
\affiliation{Instituto de Física y Astronomía, Universidad de Valparaíso, ave. Gran Bretaña, 1111, Casilla 5030, Valparaíso, Chile}
\affiliation{Chinese Academy of Sciences South America Center for Astronomy (CASSACA), National Astronomical Observatories, CAS, Beijing 100101, China}
\email{zhen.guo@uv.cl}


\author[0000-0001-9483-2016]{Zoltán Garai}
\affiliation{ELTE Gothard Astrophysical Observatory, H-9700 Szombathely, Szent Imre h. u. 112, Hungary}
\affiliation{Astronomical Institute, Slovak Academy of Sciences, 05960 Tatranská Lomnica, Slovakia}
\email{zgarai@ta3.sk}

\author[0000-0003-1305-3761]{R. Paul Butler}
\affiliation{Earth and Planets Laboratory, Carnegie Institution for Science, 5241 Broad Branch Road, NW, Washington, DC 20015, USA}
\email{pbutler@carnegiescience.edu}

\author[0000-0002-7201-7536]{Alexis Brandeker}
\affiliation{Department of Astronomy, Stockholm University, AlbaNova University Center, 10691 Stockholm, Sweden}
\email{alexis@astro.su.se}

\author[0009-0008-2801-5040]{Johanna K. Teske}
\affiliation{Earth and Planets Laboratory, Carnegie Institution for Science, 5241 Broad Branch Road, NW, Washington, DC 20015, USA}
\affiliation{The Observatories of the Carnegie Institution for Science, 813 Santa Barbara Street, Pasadena, CA 91101, USA}
\email{jteske@carnegiescience.edu}

\author[0000-0001-8627-9628]{Davide Gandolfi}
\affiliation{Dipartimento di Fisica, Università degli Studi di Torino, via Pietro Giuria 1, I-10125, Torino, Italy}
\email{davide.gandolfi@unito.it}

\author[0000-0002-0971-6078]{Haochuan Yu}
\affiliation{Sub-department of Astrophysics, Department of Physics, University of Oxford, Oxford OX1 3RH, UK}
\email{haochuan_yu@hotmail.com}

\author[0000-0003-3429-3836]{Nicolas Billot}
\affiliation{Observatoire astronomique de l'Université de Genève, Chemin Pegasi 51, 1290 Versoix, Switzerland}
\email{nicolas.billot@unige.ch}

\author[0000-0003-1453-0574]{Suzanne Aigrain}
\affiliation{Sub-department of Astrophysics, Department of Physics, University of Oxford, Oxford OX1 3RH, UK}
\email{suzanne.aigrain@physics.ox.ac.uk}

\author[0000-0002-2207-0750]{Michael Cretignier}
\affiliation{Sub-department of Astrophysics, Department of Physics, University of Oxford, Oxford OX1 3RH, UK}
\email{michael.cretignier@physics.ox.ac.uk}

\author[0000-0003-3603-1901]{Liang Wang} 
\affiliation{Nanjing Institute of Astronomical Optics \& Technology, Chinese Academy of Sciences, Nanjing 210042, People’s Republic of China}
\affiliation{CAS Key Laboratory of Astronomical Optical \& Technology, Nanjing Institute of Astronomical Optics \& Technology, Nanjing 210042, People’s Republic of
China}
\affiliation{University of Chinese Academy of Sciences, Beijing 100049, People’s Republic of China}
\email{liangwang@niaot.ac.cn}

\author[0009-0009-8313-1842]{Xuan Mao}
\affiliation{National Astronomical Observatories, Chinese Academy of Sciences, 100101, Beijing, China}
\affiliation{School of Astronomy and Space Science, University of Chinese Academy of Sciences, Chinese Academy of Sciences, 100049, Beijing, China}
\affiliation{INAF - Osservatorio Astronomico di Palermo, Piazza del Parlamento 1, 90134 Palermo, Italy}
\email{maoxuan@bao.ac.cn}

\author[0000-0001-8266-3024]{Wei M. Yuan}
\affiliation{National Astronomical Observatories, Chinese Academy of Sciences, 100101, Beijing, China}
\affiliation{School of Astronomy and Space Science, University of Chinese Academy of Sciences, Chinese Academy of Sciences, 100049, Beijing, China}
\email{wmy@nao.cas.cn}

\author[0000-0002-7259-5897]{Hongpeng Lu}
\affiliation{State Key Laboratory of Public Big Data and Guizhou Radio Astronomical Observatory, Guizhou University, Guiyang 550025, China}
\email{hplu@gzu.edu.cn}

\author[0009-0003-8853-4540]{Jiayin Li}
\affiliation{Department of Astronomy, Tsinghua University, Beijing 100084, China}
\email{lijiayin416@gmail.com}


\author[0000-0002-4644-8818]{Yann Alibert}
\affiliation{Center for Space and Habitability, University of Bern, Gesellschaftsstrasse 6, 3012 Bern, Switzerland}
\affiliation{Space Research and Planetary Sciences, Physics Institute, University of Bern, Gesellschaftsstrasse 6, 3012 Bern, Switzerland}
\email{yann.alibert@unibe.ch}

\author[0000-0002-7295-1951]{Ádám Boldog}
\affiliation{Konkoly Observatory, HUN-REN Research Centre for Astronomy and Earth Sciences, Konkoly Thege út 15-17., H-1121, Budapest, Hungary}
\affiliation{CSFK, MTA Centre of Excellence, Budapest, Konkoly Thege út 15-17., H-1121, Hungary}
\affiliation{HUN-REN-ELTE Exoplanet Research Group, Szent Imre h. u. 112., Szombathely, H-9700, Hungary}
\email{boldog.adam@csfk.org}

\author[0000-0002-9148-034X]{Vincent Bourrier}
\affiliation{Observatoire astronomique de l'Université de Genève, Chemin Pegasi 51, 1290 Versoix, Switzerland}
\email{vincent.bourrier@unige.ch}

\author[0000-0002-3288-0802]{Giovanni Bruno}
\affiliation{INAF, Osservatorio Astrofisico di Catania, Via S. Sofia 78, 95123, Catania, Italy}
\email{giovanni.bruno@inaf.it}

\author[0000-0002-5226-787X]{Jeffrey D. Crane}
\affiliation{The Observatories of the Carnegie Institution for Science, 813 Santa Barbara Street, Pasadena, CA 91101, USA}
\email{crane@carnegiescience.edu}

\author[0000-0002-8958-0683]{Fei Dai}
\affiliation{Institute for Astronomy, University of Hawai‘i, 2680 Woodlawn Drive, Honolulu, HI 96822, USA}
\email{fdai@hawaii.edu}

\author[0000-0001-7918-0355]{Olivier D. S. Demangeon}
\affiliation{Instituto de Astrofisica e Ciencias do Espaco, Universidade do Porto, CAUP, Rua das Estrelas, 4150-762 Porto, Portugal}
\affiliation{Departamento de Fisica e Astronomia, Faculdade de Ciencias, Universidade do Porto, Rua do Campo Alegre, 4169-007 Porto, Portugal}
\email{olivier.demangeon@astro.up.pt}

\author[0000-0002-8091-7526]{Alexis Heitzmann}
\affiliation{Observatoire astronomique de l'Université de Genève, Chemin Pegasi 51, 1290 Versoix, Switzerland}
\email{alexis.heitzmann@unige.ch}

\author[0009-0000-6461-5256]{Zhecheng Hu}
\affiliation{Department of Astronomy, Tsinghua University, Beijing 100084, China}
\email{hzc22@mails.tsinghua.edu.cn}

\author{Pradip Karmakar} 
\affiliation{Department of Mathematics, Madhyamgram High School (H.S.), Madhyamgram, Sodepur Road, Kolkata -700129, West Bengal, India.}
\email{pradipkarmakar39@gmail.com}

\author{Levente Kriskovics}
\affiliation{Konkoly Observatory, HUN-REN Research Centre for Astronomy and Earth Sciences, Konkoly Thege út 15-17., H-1121, Budapest, Hungary}
\affiliation{CSFK, MTA Centre of Excellence, Budapest, Konkoly Thege út 15-17., H-1121, Hungary}
\email{kriskovics.levente@csfk.org}

\author[0000-0001-9699-1459]{Monika Lendl}
\affiliation{Observatoire astronomique de l'Université de Genève, Chemin Pegasi 51, 1290 Versoix, Switzerland}
\email{monika.lendl@unige.ch}

\author[0000-0002-2412-5751]{He Y. Liu}
\affiliation{National Astronomical Observatories, Chinese Academy of Sciences, 100101, Beijing, China}
\email{liuheyang@nao.cas.cn}

\author[0000-0003-3794-1317]{Pierre F. L. Maxted}
\affiliation{Astrophysics Group, Lennard Jones Building, Keele University, Staffordshire, ST5 5BG, United Kingdom}
\email{p.maxted@keele.ac.uk}

\author[0000-0002-4047-4724]{Hugh P. Osborn}
\affiliation{Physikalisches Institut, Universität Bern, Gesellschaftsstrasse 6, 3012 Bern, Switzerland}
\affiliation{Inst. f. Teilchen- und Astrophysik, ETH Zürich, Wolfgang-Pauli-Strasse 27, 8093 Zürich, Switzerland}
\email{hugh.osborn@unibe.ch}

\author[0000-0003-2029-0626]{Gaetano Scandariato}
\affiliation{INAF, Osservatorio Astrofisico di Catania, Via S. Sofia 78, 95123 Catania, Italy}
\email{gaetano.scandariato@inaf.it}

\author[0000-0002-8681-6136]{Stephen A. Shectman}
\affiliation{The Observatories of the Carnegie Institution for Science, 813 Santa Barbara Street, Pasadena, CA 91101, USA}
\email{shec@carnegiescience.edu}

\author[0000-0001-9047-2965]{Sérgio G. Sousa}
\affiliation{Instituto de Astrofisica e Ciencias do Espaco, Universidade do Porto, CAUP, Rua das Estrelas, 4150-762 Porto, Portugal}
\affiliation{Departamento de Fisica e Astronomia, Faculdade de Ciencias, Universidade do Porto, Rua do Campo Alegre, 4169-007 Porto, Portugal}
\email{sergio.sousa@astro.up.pt}

\author[0000-0003-2417-7006]{Solène Ulmer-Moll}
\affiliation{Leiden Observatory, University of Leiden, Einsteinweg 55, 2333 CA Leiden, The Netherlands}
\affiliation{Space Sciences, Technologies and Astrophysics Research (STAR) Institute, Université de Liège, Allée du 6 Août 19C, 4000 Liège, Belgium}
\email{ulmer-moll@strw.leidenuniv.nl}

\author[0000-0003-3015-6455]{Mu-Tian Wang}
\affiliation{School of Astronomy and Space Science, Nanjing University, Nanjing 210023, China.}
\affiliation{Key Laboratory of Modern Astronomy and Astrophysics, Ministry of Education, Nanjing, 210023, People’s Republic of China}
\email{mutianwang97@gmail.com}

\author[0000-0001-8749-1962]{Thomas G. Wilson}
\affiliation{Department of Physics, University of Warwick, Gibbet Hill Road, Coventry CV4 7AL, United Kingdom}
\email{thomas.g.wilson@warwick.ac.uk}

\author[0000-0002-6937-9034]{Sharon X.~Wang}
\affiliation{Department of Astronomy, Tsinghua University, Beijing 100084, China}
\affiliation{Chinese Academy of Sciences South America Center for Astronomy (CASSACA), National Astronomical Observatories, CAS, Beijing 100101, China}
\email{sharonw@tsinghua.edu.cn}

\begin{abstract}

Young planets offer a unique window into the early stages of planetary evolution. AU Mic is one of the nearest (9.8 pc) pre-main sequence stars ($\sim$20 Myr), hosting two transiting Neptune-sized planets and a debris disk. Previous studies have shown that the rotation of the central star, the debris disk, and the inner planet b are all aligned, suggesting that the system has not undergone violent evolution. Here we report new Rossiter–McLaughlin (RM) measurements for both AU Mic b and c, which happened to transit back-to-back on Aug 24 and 25, 2024, using the Magellan Planet Finder Spectrograph (PFS), accompanioned with contanporaneous photometry from LCOGT and CHEOPS. We confirm the aligned orbit of AU Mic b ($\lambda_b=1^\circ \pm 12^\circ$) and finding two possible solutions for AU Mic c: we slightly favor an aligned solution ($\lambda_c=-10^\circ \pm 16^\circ$) but cannot rule out a polar solution ($\lambda_c=87^{\circ}\ ^{+36^{\circ}}_{-29^{\circ}}$). Broader considerations, including dynamical stability and transit possibility, also support the mutually aligned scenario. An unexpected stellar signal during ingress and the poor TTV predictions of AU Mic c prevent a precise constraint on its obliquity, and various attempts using chromatic spectral analyses fail to outperform simple data exclusion in mitigating the stellar contamination. Our observation highlights the importance of understanding stellar activity across multiple timescales and channels when characterizing young, active systems. A robust solution for the AU Mic architecture will require either a better understanding of stellar activity or future observations fortuitously free from strong stellar contamination.

\end{abstract}



\section{Introduction} \label{sec:intro}

The alignment between a planet's orbit and its host star's spin, known as stellar obliquity, is a key tracer of a planetary system's history as it encodes the dynamical interplay between star, disk, and planet. Observations over the past decades have revealed a wide diversity of obliquities in exoplanetary systems \citep[e.g.,][]{Albrecht2022}, but the origins of these misalignments are still under debate. For instance, while late infall \citep[e.g.][]{Bate2010} or magnetic warping \citep[e.g.,][]{Lai2011} could tilt protoplanetary disks and produce primordial misalignments within a few Myr, secular chaos \cite[e.g.,][]{Wu2011} may require hundreds of Myr to excite large obliquities. Measurements in young systems ($<100$~Myr) are essential to distinguish between such mechanisms operating on different timescales.

To date, nearly all planets younger than 100 Myr exhibit aligned orbits \cite[e.g., Figure 2 of][]{Teng2024}. These systems are unlikely to have undergone significant tidal realignment \citep{Hut1981,Ogilvie2014}. Therefore, current results suggest that most planets are primordially aligned and some are altered by post-formation mechanisms to become the mature misaligned population we observe, potentially acting over timescales of hundreds Myr. However, this sample only consists of $\sim$10 planets, precluding a firm conclusion. For instance, \cite{Barber2025} suggests that about 10 additional measurements are required to achieve statistical significance.

The AU Mic system is a uniquely interesting young planetary system. It is the only known young planetary system within 10~pc of the Sun \citep{Plavchan2020}, consisting of a $\sim$20 Myr-old M-dwarf, two Neptune-sized transiting planets ($\rm P_b \sim$8.5 days, $\rm P_c \sim$18.9 days), a debris disk, and a potential non-transiting planet d residing between planets b and c \citep{Wittrock2023_AU_Mic_TTV, Boldog2025_CHEOPS_TTV}. Beyond expanding the sample of young planets, AU Mic is vital for studying disk misalignments. Observations on protoplanetary disks reveal that disk misalignments are common \citep{Davies2019,Biddle2025}, which appears to conflict with the prevalent alignment observed in young planets. One possibility is that the orientation of the inner disk, where transiting planets form, differs from the outer disk and tends to remain aligned. TIDYE-1 (IRAS 04125+2902), hosting an aligned transiting planet and a misaligned transitional disk \citep{Barber2025}, presents an example consistent with this hypothesis. The combination of debris disk and transiting planets makes AU Mic another crucial system for investigating the alignment between inner and outer disks.

The obliquity of AU Mic b was measured and found to be consistent with zero shortly after its discovery \citep{Hirano2020_AU_Mic_b_RM,Martioli2020_AU_Mic_b_RM,Palle2020_AU_Mic_b_RM,Addison2021_AU_Mic_b_RM}. Combined with the nearly $90^\circ$ inclinations of the stellar rotation and the debris disk \citep{Hurt2023}, the angular momenta of the system's different components appear to be well-aligned. However, \cite{Yu2025_AU_Mic_c_RM} recently published RM measurements of AU Mic c using ESPRESSO, suggesting a possibly misaligned polar orbit. Due to the shallow transit depth of planet c and the significant stellar activity of AU Mic, their results have a large uncertainty. More recently, CHEOPS TTV results \citep{Boldog2025_CHEOPS_TTV} found a solution that favors a retrograde orbit for AU Mic c, making the system's configuration even more puzzling. Independent measurements are essential to unveil the true architecture of this interesting young system.

In this work, we conducted Rossiter-McLaughlin \citep[RM;][]{Rossiter1924,McLaughlin1924} observations during a unique window when AU Mic b and c had observable transit windows on consecutive nights (Aug.~24--25, 2024). This once-a-decade opportunity enables a direct comparison of their RM signals under similar stellar magnetic activities using the same instrument. The paper is organized as follows: In Section \ref{sec:obs}, we describe the observations and data used in this work. We introduce the main types of stellar activity signals in our data and describe attempts to mitigate them in Section \ref{sec:st_act}. Sections \ref{sec:transit} and \ref{sec:RM} present the modeling of the light curves and radial velocities (RVs), respectively. We discuss the comparison with previous studies and theoretical considerations of AU Mic c's orbit in Section \ref{sec:discuss}, and summarize our main results in Section \ref{sec:summary}.

\section{Observations and Data Reduction} \label{sec:obs}

We observed the back-to-back transits of AU Mic b and c on Aug 24th and 25th, 2024, local time in Chile (UT Aug 25-26). These observations included photometry from the space-based CHEOPS satellite and the ground-based LCOGT network to capture the light curves, alongside radial velocity (RV) measurements from the Planet Finder Spectrograph (PFS). We also conducted a two-month, ground-based photometric campaign to monitor the long-term stellar activity of AU Mic. All data used in this paper are available in Zenodo \footnote{The data is available on Zenodo under an open-source Creative Commons Attribution license: \dataset[doi:10.5281/zenodo.18910646]{https://doi.org/10.5281/zenodo.18910646}}. Below we describe each of these observations in further detail.

\subsection{CHEOPS Photometry}

Synchronously to the spectroscopic measurements, the CHEOPS space telescope \citep{Benz2021} observed AU Mic b and c, covering a $\sim$40.7\,h time window (2024-08-24 21:39 to 2024-08-26 14:21 UTC, CHEOPS file key \texttt{CH\_PR140071\_TG002207}),
which simultaneously included transits of both AU\,Mic\,b and AU\,Mic\,c positioned near the two ends of the full visit.
The observations were acquired in imagette mode with a 5\,s integration time.
Photometry was extracted from $30$-pixel radius imagettes using the
\texttt{PIPE} PSF–photometry pipeline \citep{Brandeker2024}. The effective duty cycle of this visit was 73.1\%, resulting in 21\,413 usable frames. We then binned the data to 35 s cadence by average.

Prominent outliers are visible in the resulting light curves at the start and end of each orbit, caused by Earth-scattered light correlating with telescope roll angles. We removed these by applying visually defined roll-angle masks. This approach performed similarly to background-rate masking. The outlier-removed light curve is shown in Fig.~\ref{fig:raw_data}.

\subsection{LCOGT Photometry}
We used the Las Cumbres Observatory Global Telescope \citep[LCOGT;][]{Brown2013} network's 0.35 m telescopes \citep{Harbeck2024} to obtain: (1) photometric transit observations of AU Mic c on 25 August 2024, and (2) long-term monitoring of AU Mic’s stellar variability. All observations used the 30$'$ × 30$'$ readout mode. Raw images were calibrated using the standard \texttt{BANZAI} pipeline \citep{McCully2018}, and photometry was extracted with \texttt{AstroImageJ} \citep{Collins2017}.

\subsubsection{AU Mic c transit observations}
The transit of AU Mic c on 25 August 2024 was observed from Cerro Tololo Inter-American Observatory (CTIO) in Chile. Two telescopes simultaneously obtained 30-s exposures in the Sloan $i'$ filter over $\sim6.4$~h. These observations were planned using a customized version of the \texttt{TAPIR} software \citep{Jensen2013}.

\subsubsection{Long-term stellar variability monitoring}
We monitored AU Mic's long-term variability from 25 August to 25 October 2024, using multiple LCOGT 0.35 m telescopes at CTIO, Haleakalā Observatory, Siding Spring Observatory, and the South African Astronomical Observatory. We acquired frequent imaging sequences, typically consisting of 10 images per filter, with per-image exposure times of $\leq30$~s. In total, 118, 112, and 2 successful sequences were obtained in the Sloan $r'$, $i'$, and Bessel $V$ filters, respectively. For each sequence (set of images) and filter, photometric measurements were median-combined, with standard deviations adopted as uncertainties.

\subsection{Planet Finder Spectrograph (PFS)} \label{subsec:PFS_obs}

We observed AU Mic on the nights of August 24 and 25, 2024 (Chile time) using the Planet Finder Spectrograph (PFS, \citealt{Crane2006_PFS1,Crane2008_PFS2,Crane2010_PFS3}) on the Magellan II Clay 6.5 m telescope at Las Campanas Observatory, covering the transits of AU Mic b and c, respectively. We obtained 37 and 45 exposures, utilizing a 600s integration time and a $1\times2$ binning readout (readout time: $\sim$62s). Observations were conducted with the iodine cell and a 0.3\arcsec $\times$2.5\arcsec\ slit, providing a resolving power of $R = 127\,000$.

\begin{figure*}
\centering
\includegraphics[width=\textwidth]{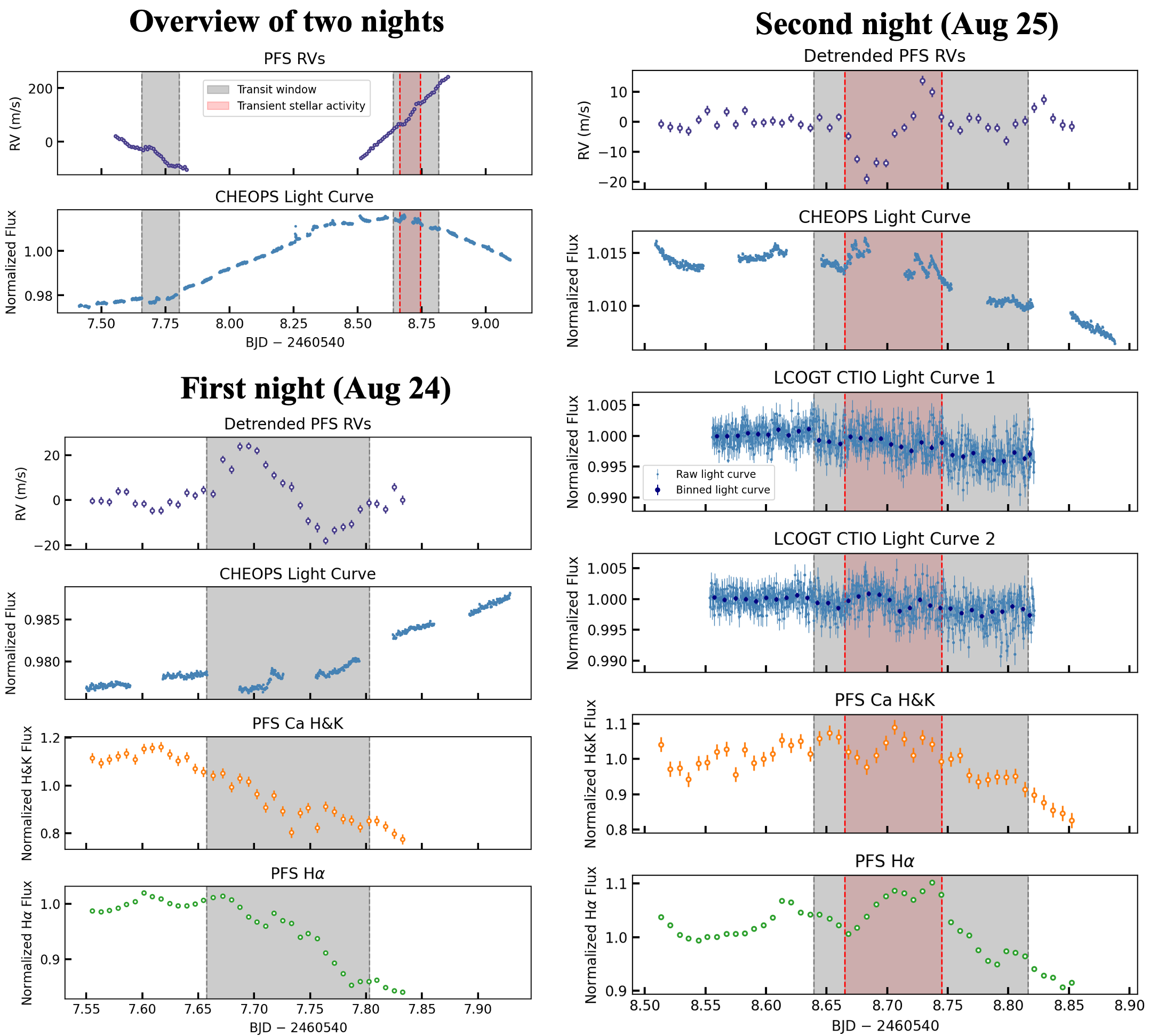}
\caption{RVs, light curves, and stellar activity indicators for AU Mic b and c transits observed on consecutive nights (local dates labeled). Gray shaded regions denote the transit window for each planet. Red shaded regions highlight a transient stellar activity event on the second night that affected all collected data during planet c's transit.} 
\label{fig:raw_data}
\end{figure*}

The spectral data were reduced and RVs computed using a customized pipeline following \citet{Butler1996}. The median uncertainty of the final RVs is 1.7 m/s. 
Additionally, we extracted two common stellar activity indicators: the emission fluxes of the Ca II H\&K lines and the $\rm H\alpha$ line. Flux time series were derived following the methodology of \citet{Perdelwitz2021_SHK}, with slight adjustments to the wavelength bands to account for AU Mic's strong emission. The resulting time series are shown in Fig.~\ref{fig:raw_data}.

\section{Modeling stellar signals} \label{sec:st_act}

Stellar activity is evident in both the light curves and RV data (Fig.~\ref{fig:raw_data}). We classify these signals into two categories: long-term variations (timescale $> 1$ day) and short-term transient variations (timescale $\sim$ several hours). We describe our modeling of these two variations in the following subsections.

\subsection{Long-term variation: spots} \label{subsec:st_rot}

The observed light curves and RVs both exhibit large-amplitude long-term variations ($>1 \%$ in light curves and $>100$ m/s in RVs). These variations are consistent with previous TESS light curves and other RV observations, most likely caused by rotationally modulated starspots \citep[e.g.,][L. Kriskovics et al. in prep.]{Plavchan2020, Szabo2022}. We confirmed this by applying a simple spot model using \texttt{starry} \citep{Luger2019_starry}, which successfully reproduced both datasets. However, for baseline detrending, this physical model does not outperform a low-order polynomial. We detailed the spot model and our justification for adopting low-order polynomials in Appendix \ref{sec_app:spot}. For the subsequent analysis of the light curves and RVs, we adopted a nightly polynomial to model the long-term trends.

\subsection{Transient variation: flares?} \label{subsec:st_transient}

The CHEOPS light curves exhibit numerous visible flare-like signals. While most do not impact our analysis, a set of flare-like variations near or during the predicted transit of planet c significantly obscures the transit signal (red region in Fig.~\ref{fig:raw_data}). Concurrently, an RV variation is present and resembles the RM signal of a retrograde planet: an initial blueshift followed by a redshift, with a peak-to-peak amplitude of $\sim 30$ m/s. However, we can rule out an RM-effect origin, as their duration ($\sim 2$ hours) is significantly shorter than planet c's $\sim 4$-hour transit \citep{Wittrock2023_AU_Mic_TTV}. We also rule out spot-crossing since the flux exceeds the out-of-transit baseline. As the events in the light curves and RVs are temporally aligned, we posit that they originate from the same stellar activity phenomenon.

To investigate and mitigate this, we performed a chromatic RV analysis (detailed in Appendix \ref{sec_app:chromatic_rv}). Notably, Figure \ref{fig:rv_chunk} highlights that in 10 specific sub-wavelength ranges (chunks), the RV variations maintain the morphology but exhibit substantially larger amplitudes. We find that 8 of these 10 ranges match activity-sensitive lines identified by \cite{Wise2018}, predominantly Mg I and Fe I lines. This may unveil a chromospheric transient origin. Further discussion regarding the short-term activity in our dataset is provided in Appendix \ref{sec_app:variability}.

However, attempts using chromatic RV analysis still fail to remove these anomalous variations cleanly. In the subsequent analysis, we choose to mask out the affected time range in both the light curves and RVs.

\begin{figure}
\centering
\includegraphics[width=0.49\textwidth]{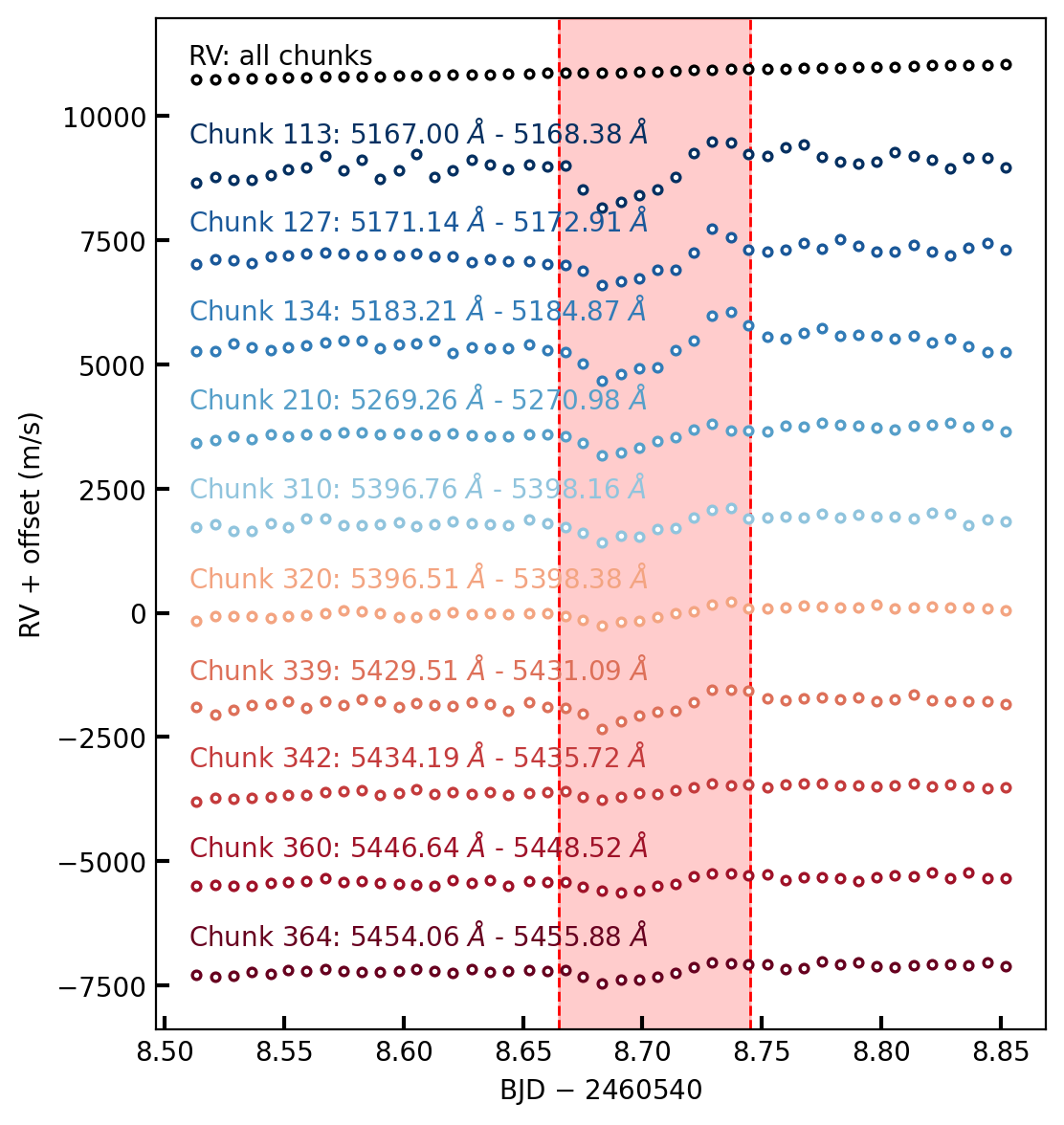}
\caption{RVs of the 10 PFS spectral chunks most affected by the transient stellar activity, plotted alongside the combined RVs from all chunks for comparison. Wavelength ranges are indicated in the plot for each chunk (star rest frame in air). The red regions denote the duration of the transient stellar activity. Note that the amplitude of activity in these chunks reaches thousands of m/s. On this scale, the influence on the combined RVs is not visible.} 
\label{fig:rv_chunk}
\end{figure}

\section{Transit fitting} \label{sec:transit}

\begin{figure*}
\centering
\includegraphics[width=0.95\textwidth]{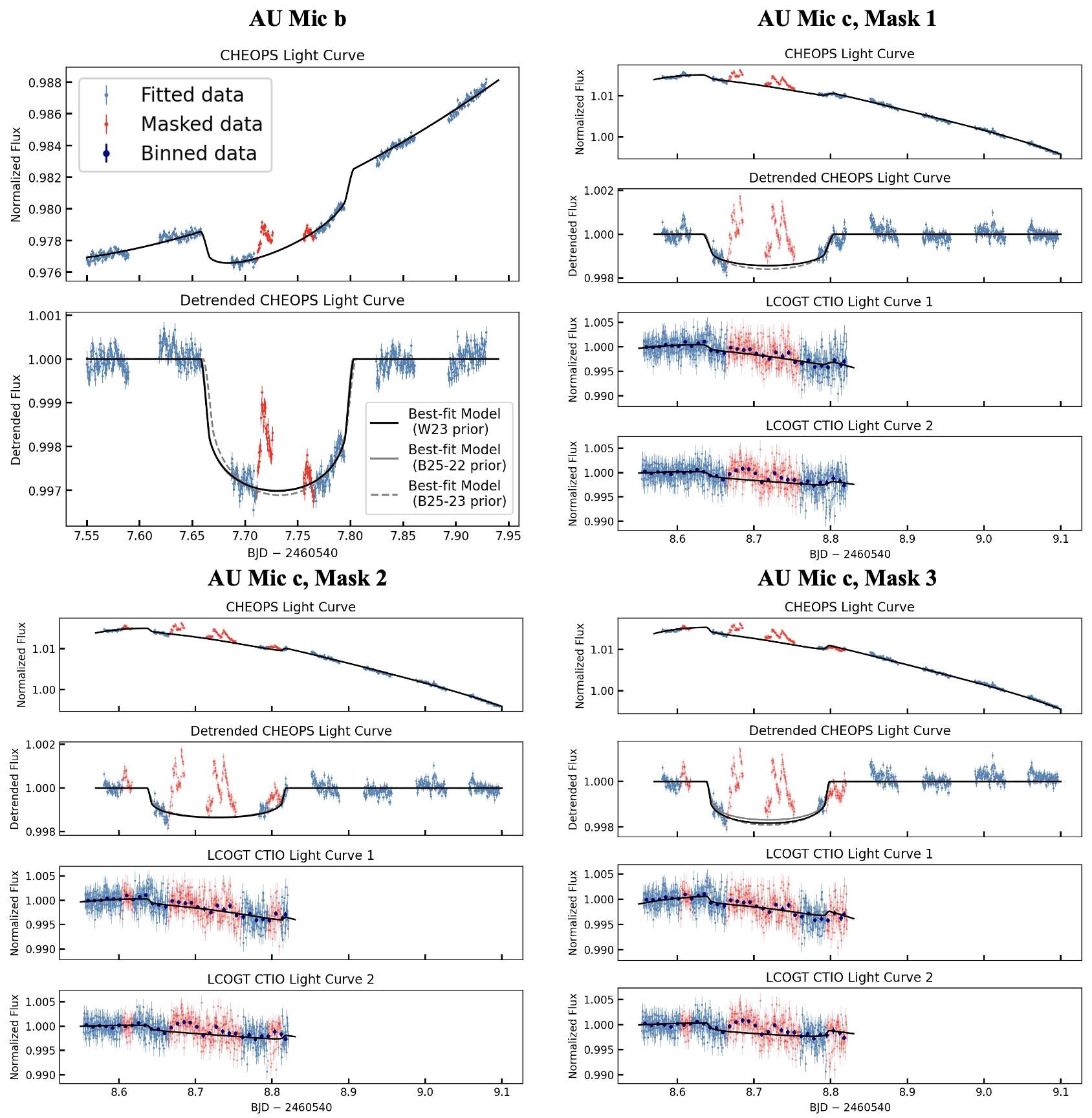}
\caption{Transit fitting results for AU Mic b and c. For AU Mic c, three fits using different flare-masking choices are presented. Blue dots represent the data used in the fit, while red dots indicate masked data points. The solid black lines represent the best-fit transit models using the W23 prior. For the detrended CHEOPS data, additional best-fit models using alternative priors are shown as solid and dashed gray lines; these exhibit only marginal deviations from each other. \textit{Priors used:} W23: based on \cite{Wittrock2023_AU_Mic_TTV}, B25-22 and B25-23: based on 2022 and 2023 values of \cite{Boldog2025_CHEOPS_TTV}, respectively.}
\label{fig:transit}
\end{figure*}

The primary goal of this section is to constrain the mid-transit time ($T_0$) by fitting the light curves, thereby providing critical priors for the subsequent RM modeling. We chose not to perform a joint fit combining the light curves with RVs because different priors and flare-masking strategies in the photometry yielded differences in $T_0$ that exceeded the individual statistical errors. Consequently, we decided to pass $T_0$ to the RM modeling with manually inflated error bars to account for these systematic uncertainties.

To minimize the impact of stellar activity, we modeled individual transits using local baselines rather than the complete CHEOPS dataset. We utilized the Python package \texttt{batman} \citep{Kreidberg2015_batman} for transit models and \texttt{dynesty} \citep{Speagle2020_dynesty1, Koposov2024_dynesty2} for dynamical nested sampling. We adopted low-order polynomial baselines following arguments in Section \ref{subsec:st_rot}. For planet c, we modeled the trends separately across its three datasets, as the trends for ground-based observations may be a combination of stellar rotation and atmospheric effects. Based on Bayesian evidence ($\log \mathcal{Z}$), we selected a 2nd-order polynomial baseline for planet b and two ground-based datasets for planet c, and a 4th-order polynomial for the CHEOPS data for planet c.

As these polynomials are local approximations of the spot behavior, the need for different orders in different light curve segments is expected. The higher-order polynomial required for c likely arises because the baseline is longer and the shape around the photometric peak is asymmetric. The 4th-order polynomial maintains a convex shape without exhibiting signs of overfitting and is thus preferred for our fit.

Because we fitted single transits with incomplete coverage, we applied informative transit priors derived from previous studies. We further added a normal prior on the transit duration ($T_{14}$) when calculating the model likelihood. However, transit parameters vary significantly across publications \citep[e.g.,][]{Mallorquin2024}. We tested three sets of priors: one based on \cite{Wittrock2023_AU_Mic_TTV} (hereafter W23) and two based on \cite{Boldog2025_CHEOPS_TTV} (hereafter B25-22 and B25-23, corresponding to their 2022 and 2023 values, respectively). The former study fitted transits using TESS data and extensive ground-based photometry before 2022, allowing for free eccentricity. The latter study utilized 2022 and 2023 CHEOPS data and assumed circular orbits for both planets. Consequently, these priors primarily differ in eccentricity ($e$), planet-star radius ratio ($R_p/R_s$), argument of periapsis ($\omega$), impact parameter ($b$), quadratic limb-darkening coefficients ($u_1$, $u_2$), and $T_{\rm 14}$.

We adopted uniform priors for the mid-transit time ($T_0$) and polynomial coefficients, along with a broad log-uniform prior for the jitter term. A summary of all priors is provided in Appendix Table \ref{tab:transit}.

We visually masked stellar flares (or potential spot crossing) in fitted light curves. However, the egress in the CHEOPS data near $\text{BJD} \sim 2460548.8$ is ambiguous, so we tested three different masking strategies around this time range (Fig. \ref{fig:transit}).

As a result, different priors shift $T_{0,b}$ between 2460547.730 and 2460547.733, with a typical statistical error of 0.0008 days. For AU Mic c, differences arising from the choice of priors are smaller than the typical statistical error ($\sim 0.001$ days). However, the choice of masking strategy introduces larger discrepancies: Mask 2 yields $T_{0,c} \approx 2460548.727$, while Masks 1 and 3 yield $T_{0,c} \approx 2460548.719$ (Fig.~\ref{fig:transit}). For the subsequent RM modeling, we adopted $T_{0,b} = 2460547.731 \pm 0.005$ and $T_{0,c} = 2460548.723 \pm 0.01$.

Predicting $T_0$ from previous transit observations is complicated by the transit-timing variations (TTVs) of AU Mic b and c. The TTV model is currently not well-constrained and is influenced by the non-transiting planet d \citep[e.g.,][]{Wittrock2023_AU_Mic_TTV, Boldog2025_CHEOPS_TTV}. Despite this, we collected the currently available transit times for AU Mic c and obtained an updated linear ephemeris. The resulting mid-transit time for our epoch, $T_{0,c} = 2460548.7278$ $\rm BJD_{TDB}$, is consistent with our fitting results. We also compared the results with additional transits observed by CHEOPS during the 2024 visibility season (Z. Garai et al., in prep.) and found overall consistency, confirming the reliability of the best-fit transit times used in this paper.

\section{RM effect modeling} \label{sec:RM}

In this section, we jointly fitted the RM effect for AU Mic b and c using PFS data. We excluded the transient stellar activity observed on the second night (Section~\ref{subsec:st_transient}). We modeled the data using the RM effect plus a polynomial trend to account for the stellar rotational activity. We tested fits 
without (fiducial) and with Gaussian processes (GP) for the short-timescale stellar variability (i.e., not rotation).

\subsection{Fiducial fitting of PFS RVs} \label{subsec:RM_Fi}

In this fiducial model, we used the Python package \texttt{rmfit} \citep{Stefansson2022_rmfit} for the RM model, which is based on the model by \cite{Hirano2011}. For the stellar rotational trend, we applied separate, low-order polynomials for each of the two nights. We evaluated the optimal polynomial orders by comparing their Bayesian evidence, $\log \mathcal{Z}$. Consequently, we adopted a 3rd-order polynomial for the first night (planet b) and a 2nd-order polynomial for the second night (planet c). We performed nested sampling using \texttt{dynesty} with 1000 live points.

We set Gaussian priors for the mid-transit time $T_0$ as described in Section \ref{sec:transit}. Because remaining orbital parameter posteriors from light curve fitting are highly sensitive to prior choices, we did not use them as priors. Instead, we tested the same three sets of priors used in the light curve analysis. We assigned uniform priors to the projected rotational velocity ($v \sin i$), projected stellar obliquity ($\lambda$), and polynomial trend coefficients. A complete summary of our prior settings is provided in Appendix Table \ref{tab:RM}.

\begin{figure*}[t]
\centering
\includegraphics[width=0.95\textwidth]{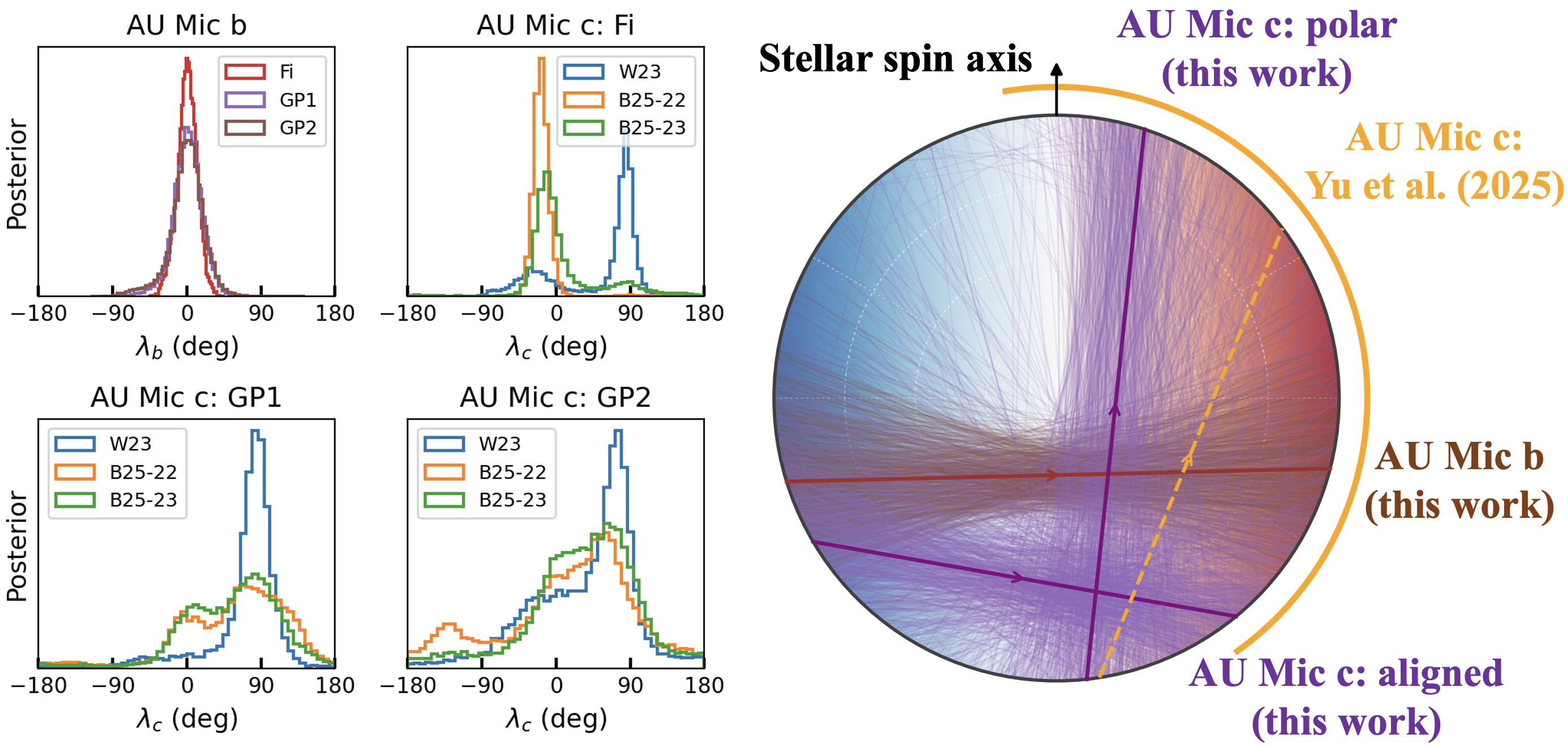}
\caption{\textit{Left panels}: Posterior distributions of the projected stellar obliquities of AU Mic b and c ($\lambda_b$ and $\lambda_c$) for representative modeling results, presenting differences in prior choices and the inclusion of the GP. \textit{Right panel}: Illustration of the transit configurations of AU Mic b and c derived from our posteriors. We plot the fiducial model under the B25-22 prior for b, and the fiducial model under the B25-23 and W23 priors simultaneously for c. Thin lines represent randomly selected samples from the posterior samples. Bold solid lines denote the median for each solution. The orange dashed line and solid arc show the median model and 2$\sigma$ range constrained by \cite{Yu2025_AU_Mic_c_RM}, respectively. The background blue and red colors indicate local RVs (blueshift and redshift) across the stellar hemispheres induced by stellar rotation. \textit{Abbreviations used in this plot:} Fi: fiducial model; GP1: GP model with a timescale prior from \cite{Yu2025_AU_Mic_c_RM} ($\log \rho_{\rm GP} \sim -4$); GP2: GP model with a timescale prior trained on activity indicators ($\log \rho_{\rm GP} \sim -2.8$). W23: transit prior based on \cite{Wittrock2023_AU_Mic_TTV}, B25-22 and B25-23: transit prior based on 2022 and 2023 values of \cite{Boldog2025_CHEOPS_TTV}, respectively.}
\label{fig:posterior}
\end{figure*}

The key results across these tests are presented in Table \ref{tab:fi} and Fig.~\ref{fig:posterior}. We summarize key features below:

\begin{enumerate}
    \item AU Mic b's obliquity ($\lambda_b$) consistently yields a median near $0^\circ$ with an uncertainty around $15^\circ$. This is consistent with previous measurements \citep{Hirano2020_AU_Mic_b_RM,Martioli2020_AU_Mic_b_RM,Palle2020_AU_Mic_b_RM,Addison2021_AU_Mic_b_RM}.
    \item Two primary families of solutions emerge for AU Mic c's obliquity ($\lambda_c$) in all tests (Fig.~\ref{fig:posterior}): an ``aligned solution" near $\lambda_c=0^\circ$ and a ``polar solution" near $\lambda_c=90^\circ$.
    \item The aligned solution favors a shorter transit duration ($T_{14,c}$) and larger impact parameter ($b_c$) for planet c, while the polar solution favors the opposite (Table \ref{tab:fi}). The posteriors for the remaining parameters are nearly identical between the two solutions.
    \item Consistent with these correlations, the occurrence and relative preference of these two families depend primarily on the $T_{14,c}$ and $b_c$ priors. For instance, the W23 prior has a smaller $b_c$ and a larger $T_{14,c}$ and favors the aligned solution.
\end{enumerate}

\begin{table*}
    \centering
    \caption{Priors and posteriors from fiducial models for the projected stellar obliquities of AU Mic b and c ($\lambda_b$ and $\lambda_c$), as well as the impact parameter ($b_c$) and transit duration ($T_{14,c}$) of c. Two families of solutions emerge in the $\lambda_c$ space (Fig.~\ref{fig:posterior}). We bifurcate the sample using a visually adopted threshold of $40^\circ$ and report the 68.3\% confidence intervals (CIs) for each sub-sample in this table. For priors, $\mathcal{N}$($\mu$, $\sigma$) means a normal prior with mean $\mu$ and standard deviation $\sigma$. $\mathcal{U}$(a, b) stands for a uniform prior ranging from a to b. $\mathcal{TN}$($\mu$, $\sigma$, a, b) stands for a truncated normal prior with mean $\mu$, standard deviation $\sigma$, lower bound $a$, and upper bound b. The W23 prior is based on \cite{Wittrock2023_AU_Mic_TTV}, and the B25-22 and B25-23 priors are based on 2022 and 2023 values of \cite{Boldog2025_CHEOPS_TTV}, respectively. Notably, the aligned solution shows a slight preference for a shorter $T_{14,c}$ and a larger $b_c$ (and vice versa for the polar solution). Consequently, the relative preferences of these two solutions shift following the chosen priors.}
    \begin{tabular}{lcccc}
    \hline\hline
     & $\lambda_b$ & $\lambda_c$ & $b_c$ & $T_{\rm 14,c}$\\
    \hline
    Fiducial, W23 prior\\
    Prior & $\mathcal{U}$(-180, 180) & $\mathcal{U}$(-180, 180) & $\mathcal{TN}$(0.3, 0.2, 0, 1) & $\mathcal{N}$(0.1765,0.01)  \\
    Aligned solution & $4 \pm 15$ & $-25 \pm 28$ & $0.36_{-0.19}^{+0.14}$ & $0.169 \pm 0.008$  \\ 
    Polar solution (slightly favored) & $6 \pm 17$ & $84 \pm 10$ & $0.21_{-0.11}^{+0.15}$ & $0.179 \pm 0.009$  \\ 
    \hline
    Fiducial, B25-22 prior$^{[1]}$\\
    Prior & $\mathcal{U}$(-180, 180) & $\mathcal{U}$(-180, 180) & $\mathcal{TN}$(0.61, 0.12, 0, 1) & $\mathcal{N}$(0.1625,0.013)  \\
    Aligned solution & $1 \pm 12$ & $-18 \pm 9$ & $0.66 \pm 0.05$ & $0.156 \pm 0.007$  \\
    \hline
    Fiducial, B25-23 prior\\
    Prior & $\mathcal{U}$(-180, 180) & $\mathcal{U}$(-180, 180) & $\mathcal{TN}$(0.66, 0.2, 0, 1) & $\mathcal{N}$(0.1554,0.027)  \\
    Aligned solution  (slightly favored) & $-2 \pm 10$ & $-10 \pm 16$ & $0.65 \pm 0.08$ & $0.154 \pm 0.015$ \\
    Polar solution & $-3 \pm 11$ & $87_{-29}^{+36}$ & $0.59_{-0.20}^{+0.12}$ & $0.163 \pm 0.025$ \\
    \hline
    \end{tabular}
    \begin{tablenotes}
    \item[1]  [1] There is nearly no polar solution for this model, so we only report the 68.3\% CIs for the whole posteriors.
    \end{tablenotes}
    \label{tab:fi}
\end{table*}

\subsection{Fitting with Gaussian process} \label{subsec:RM_GP}

We first evaluated the need to add a GP component. To do so, we utilized the structure function (SF) \citep{Simonetti1985} to analyze the correlated noise in the RV residuals after subtracting the best-fit model in Section \ref{subsec:RM_Fi}, as well as the Ca H\&K and H-$\alpha$ time series. Briefly, the SF is a time-domain diagnostic tool that calculates data variance as a function of time lag ($\Delta t$). A distinct peak at a specific time scale would signify the presence of correlated noise rather than pure white noise. To assess the statistical significance of any observed peaks, we established confidence thresholds using permutation tests. We found no significant peaks in either the residuals or the activity indicators for time scales shorter than 0.5 days. Therefore, we conclude that correlated noise in our data, aside from the rotation signals, is insignificant. Nevertheless, we still tested modeling with GPs to be consistent with \cite{Yu2025_AU_Mic_c_RM} and to check the impact of potential small-timescale stellar jitter, though small. We implemented GP models using a Mat\'ern 3/2 kernel via the Python package \texttt{celerite} \citep{Foreman-Mackey2017_celerite}. The remaining model settings were identical to those in the fiducial fit.

We set a uniform prior on GP amplitude ($\log \sigma$) and tested three timescale ($\log \rho$) priors. First, we adopted the same prior used in \cite{Yu2025_AU_Mic_c_RM} ($\log \rho \sim -4$). Second, we trained a separate GP on the Ca H\&K and H-$\alpha$ flux time series and then used the resulting $\log \rho$ posteriors as priors for the RM fit ($\log \rho \sim -2.8$). Furthermore, we replaced the 2nd-order polynomial for c with a linear trend, as this is now statistically preferred based on $\log \mathcal{Z}$.

Compared to the fiducial models, the GP models introduce the following key differences:
\begin{enumerate}
    \item Across all prior sets, the GP models are generally favored over their corresponding fiducial models by $\Delta \log \mathcal{Z} \sim 4$.
    \item The uncertainty in $\lambda_b$ is maintained or only marginally increased.
    \item The two families of solutions flatten and appear to merge into a single, broad solution (Fig.~\ref{fig:posterior}). Generally, the GP increases the statistical preference for the polar solution.
\end{enumerate}

Finally, we attempted to apply the GP to the full RV dataset, including the RVs affected by the transient activity. In this case, the required GP amplitude becomes comparable with the RM signal, leaving the posteriors for $\lambda_b$ and $\lambda_c$ largely unconstrained. We therefore conclude that our current GP implementation cannot adequately model this transient activity.

\subsection{Model comparison}

In Section \ref{subsec:RM_Fi} and \ref{subsec:RM_GP}, we find two potential solutions for $\lambda_c$: one near $\lambda_c = 0^\circ$ and another near $\lambda_c = 90^\circ$. In this section, we detail our model comparisons and present our preferred solution, though we caution that our preferences are not a definitive resolution for the architecture of AU Mic c.

\begin{figure*}
\centering
\includegraphics[width=0.95\textwidth]{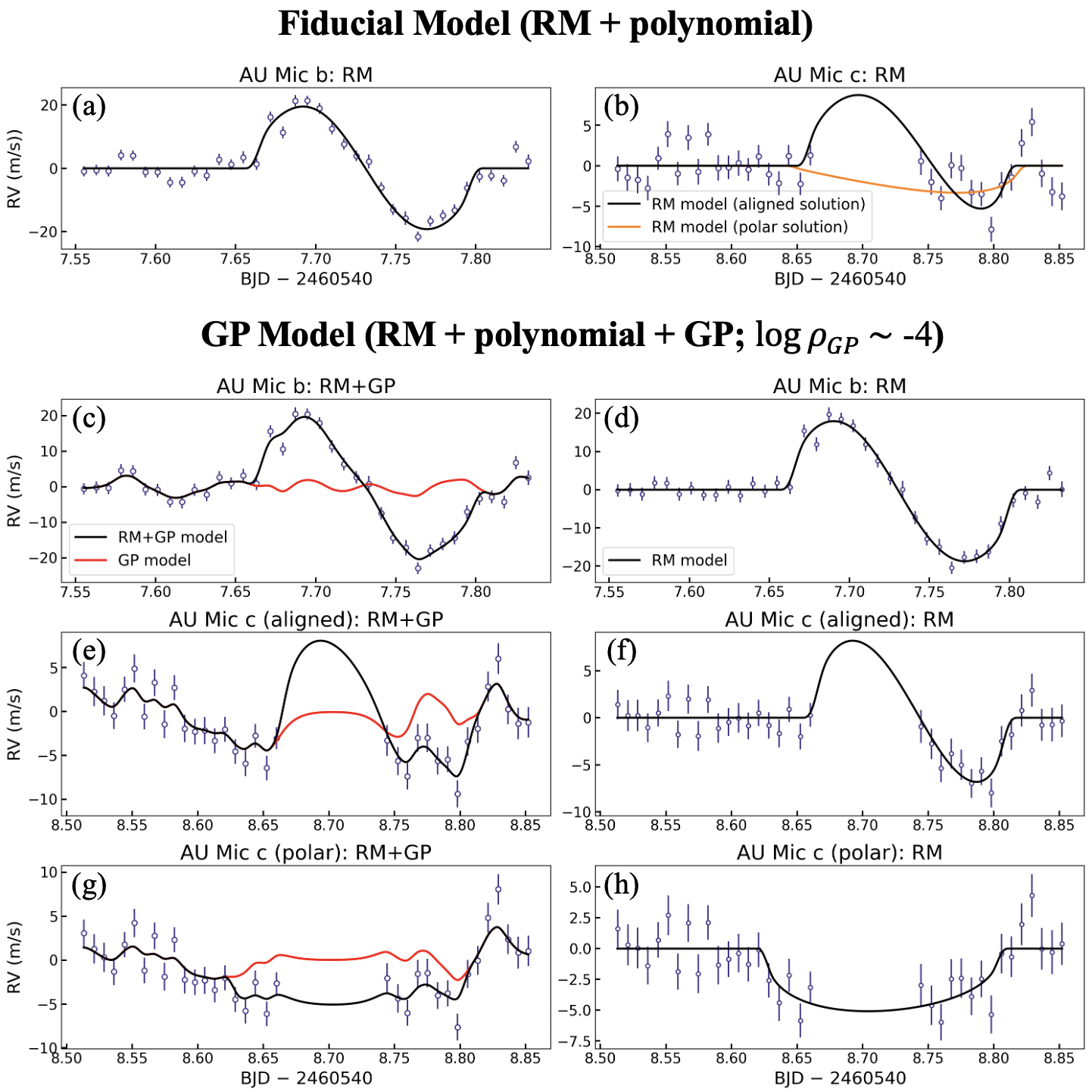}
\caption{Representative modeling results of the RM effect for AU Mic b and c. The best-fit models shown here for b and the aligned solution for c utilize the B25-22 priors (2022 values from \citealt{Boldog2025_CHEOPS_TTV}), and the polar solution for c uses the W23 priors \citep{Wittrock2023_AU_Mic_TTV}. Additional model plots are provided in Appendix \ref{sec_app:RM}. \textit{Top row}: PFS RVs and partial models after subtracting low-order polynomial trends for the fiducial models. Two solutions for c are distinguished by color. \textit{Bottom three rows}: Fitting results for models incorporating GP, again showing two solutions for AU Mic c. In each of these rows, the left panel displays PFS RVs and the model with the polynomial trend subtracted, leaving the combined RM and GP components. The right panel isolates the RM signal by further subtracting the GP component. The GP fit shown here adopts the time-scale prior ($\log \rho$) from \cite{Yu2025_AU_Mic_c_RM}.} 
\label{fig:RM}
\end{figure*}

We first evaluate the GP models. Adding a GP component is adding significant complexity to the model. For the RM model of AU Mic c, a notable discrepancy occurs at the $\sim$4 data points immediately before the data gap. Without a GP, the $\sim$4 data points are constrained near the baseline (Fig.~\ref{fig:RM} b), which results in preferences toward short-transit-duration solutions. Incorporating the GP allows these $\sim$4 points to be modeled as a combined GP+RM signal, which enables solutions with longer transit durations (Fig.~\ref{fig:RM} g,h). As discussed in Section \ref{subsec:RM_Fi}, polar models favor longer transit durations. This explains why the GP increases the preference for the polar solution, particularly in the W23 priors, which already include a prior for long transit durations. However, it is unclear whether attributing these $\sim$4 points to a combined GP+RM model represents a physical signal or overfitting. This highlights a risk for GP to “absorb” (thus change) the RM signal, which calls for further assessment of robustness.

To assess the robustness of the GP models versus the fiducial model, we performed injection-recovery tests on AU Mic c's RM signal. We generated RVs following various ``true configurations'' of different $\rm \lambda_c$ values and added RV noise using the residuals from AU Mic b's fit. We chose to use planet b's residuals because our fit has probably captured the true RM signals of planet b, which is well-established from previous studies \citep[e.g.,][]{Palle2020_AU_Mic_b_RM}, and thus the residuals are most likely close to the true noise structure, including photon noise and instrumental and stellar jitter.

Our tests indicate that the fiducial model successfully recovers the injected signals, although the extracted $\rm \lambda_c$ can sometimes be biased when incorrect priors were used. Adding a GP always improves the $\log \mathcal{Z}$ by $\sim$4, likely because it better fits the short-term variations in the baseline, plausibly a couple of mini-flares (e.g., near BJD$\sim$2460547.60). The GP also tends to yield more conservative results with larger uncertainties for $\rm \lambda_c$, making it appear less sensitive to incorrect priors. However, in some tests, even when provided with correct priors and achieving a higher $\log \mathcal{Z}$, the GP models report biased results away from the true values. Therefore, while a GP yields a statistically better fit, it does not guarantee a more robust extraction of the RM signal.

Our injection-recovery tests demonstrate that our fiducial model provides accurate results when priors are handled carefully. The consistent results between the fiducial and GP models for planet b further support that the simpler, fiducial model should work just as well. Given the risk and problem with GP, we adopt the fiducial model without GP as our preferred solution. Regarding the risk of bias by incorrect priors, we tested different sets of priors in Section \ref{subsec:RM_Fi} and discuss our preference below. We also highlight that more robust transit parameters from future studies will help improve our constraints.

Next, we discuss the relative preference of the two solutions. Beyond $\lambda_c$, the two solutions diverge primarily in $b_c$ and $T_{14,c}$. When further comparing the literature, we find that models assuming circular orbits during light curve fitting tend to derive a larger $b_c$ and a shorter $T_{14,c}$ than free-eccentricity models (for a direct comparison, see \citealt{Mallorquin2024}). Consequently, we broadly conclude that assuming a circular orbit drives a stronger preference for the aligned solution in our data.

Although the eccentricities of AU Mic b and c remain a subject of ongoing debate, current TTV analyses support a circular orbit \citep{Wittrock2023_AU_Mic_TTV,Boldog2025_CHEOPS_TTV}. Based on this evidence, we suggest the aligned solution, which implies that AU Mic b and c are coplanar, an architecture consistent with the majority of known multi-transiting systems \citep{Winn2015}. 

We choose to report the conservative results (those with larger errors on $\lambda$) from B25-22 and B25-23. As a conclusion, we obtain $\lambda_b = 1^\circ \pm 12^\circ$. For $\lambda_c$, we slightly favor the aligned solution of $\lambda_c=-10^\circ \pm 16^\circ$, but cannot completely rule out the polar solution of $\lambda_c={87^\circ}_{-29^\circ}^{+36^\circ}$.

\section{Discussion} \label{sec:discuss}

\subsection{Comparison with ESPRESSO results}

Recently, \cite{Yu2025_AU_Mic_c_RM} published independent ESPRESSO RM observations of AU Mic c using ESPRESSO on Jul 7 and 26, 2023. There are several differences between the datasets and analyses. First, although their and our datasets were both significantly affected by stellar activity, the morphology differed. For instance, the largest RV variation in the ESPRESSO data coincided with a typical $\sim$1\% flare in the photometry, whereas ours aligned with a smaller ($\sim$0.1\%), more complex brightness fluctuation.

Second, due to the distinct wavelength coverages and calibration methods of PFS and ESPRESSO, our activity mitigation strategies necessarily differed. While \cite{Yu2025_AU_Mic_c_RM} performed a dedicated line-by-line analysis and effectively removed a significant portion of the flare signals, we found that a similar approach was less effective for our dataset, likely due to the limitations inherent to iodine cell spectrographs or the different nature of the activity observed. However, the RV derivation of PFS data performed a weighted average over wavelength segments, thereby mitigating stellar activity to some degree. 

Finally, \cite{Yu2025_AU_Mic_c_RM} performed a joint fit of RVs and light curves. We opted against a joint fit because the derived mid-transit time in our observations is highly sensitive to the choice of flare masks. 

Despite these methodological and observational differences, both studies yield similar posteriors. \cite{Yu2025_AU_Mic_c_RM} reported $\lambda_c = 67.8^{\circ+31.7^{\circ}}_{-49.0^{\circ}}$, which is consistent with both of our aligned and polar solutions. Incorporating GP in our fit brings the overall $\lambda_c$ posteriors even closer together. However, how this similarity reflects the true obliquity of AU Mic c remains unclear.

We attempted to jointly fit our RVs with those of \cite{Yu2025_AU_Mic_c_RM} using our fiducial settings. However, we found that the ESPRESSO results were highly sensitive to the $T_{0,c}$ priors (particularly for their second transit) when a joint photometry-RV fit is not performed. Applying a tight $T_{0,c}$ prior around the best-fit value from \cite{Yu2025_AU_Mic_c_RM} drived the solution toward $\lambda_c \approx 90^{\circ}$, whereas a loose prior favored $\lambda_c \approx 0^{\circ}$. Additionally, the joint fit of ESPRESSO and PFS data would be dominated by the ESPRESSO data due to its higher cadence and RV precision.

Future works combining more ground-based and CHEOPS photometry with more transit epochs, in tandem with photodynamic modeling, would shine more light on the exact transit ephemerides of these RM observations, potentially enabling better constraints on the obliquities and mutual inclinations of planet b \& c, but this is beyond the scope of this paper.

\subsection{Challenges of mutually inclined orbits of AU Mic b and c} \label{subsec:dis_obl}

While our results and \cite{Yu2025_AU_Mic_c_RM} cannot rule out the solution of $\lambda_c \sim 90^\circ$, we discuss several considerations that challenge this scenario. The first consideration is the dynamical stability. High mutual inclinations could activate secular and high-order resonances that are suppressed in coplanar systems. This increases the chance of orbit overlap, which triggers chaos and reduces stability. Given its tightly packed orbits and near-commensurable period ratios, AU Mic is already near the edge of stability. \cite{Hu2025_stability} showed that the system becomes unstable in a coplanar configuration with $e \sim 0.04$ when AU Mic d is included. Furthermore, \cite{Yu2025_AU_Mic_c_RM} found the b-c system becomes unstable at mutual inclinations $\gtrsim 50^\circ$.

Another critical consideration is the geometric transit probability. For AU Mic b and c, the joint transit probability is $\sim$3\% if the system is coplanar, but decreases to $\sim$0.2\% if the orbits are mutually perpendicular. Furthermore, \cite{Teng2025} notes that the time-averaged probability of a double transit decreases even further when accounting for nodal precession induced by mutual inclination. The precession might also lead to observable transit duration variations with the change rate of the impact parameter in the range of 0.1--1 per 10 years, according to our simple N-body simulation estimates. This provides a potential testable parameter for future observations.

\subsection{Population-level theoretical expectations for the obliquity of AU Mic c} \label{subsec:dis_form}

In this section, we contextualize the obliquity of AU Mic c within our current understanding of planet formation and evolution. For multi-planetary systems, \cite{Dai2024} demonstrated that the fraction of neighboring planet pairs near a first-order mean-motion resonance is $\sim$70\% for systems younger than 100 Myr, declining to $\sim$15\% for mature systems. Furthermore, most young multi-planetary systems (and even some single-planet systems) exhibit TTVs \citep{Lopez_Murillo2026}. Additionally, as discussed in Section \ref{sec:intro}, nearly all young planets ($<$100 Myr) in the current observation sample are aligned \citep[e.g.,][]{Teng2024,Barber2025}. Together, these findings provide direct evidence that close-in planets undergo Type I migration and resonance capture within the protoplanetary disk \citep[e.g.,][]{Izidoro2017}.

Aside from the obliquity of AU Mic c, the properties of the AU Mic system fit these trends well: the two transiting planets are near-resonant, exhibit TTVs, and AU Mic b is aligned. Therefore, we expect that the AU Mic planets formed via this same standard pathway, or at least are coplanar, similar to other multi-planetary systems \citep{Fabrycky2014}. Conversely, if AU Mic c is misaligned, it would require a post-formation mechanism capable of driving planet c to a polar orbit while simultaneously leaving planet b aligned, preserving the near-resonant period ratio, maintaining detectable TTVs, and leaving no other dynamical fingerprints to distinguish AU Mic from the broader population of young systems. Under the current theoretical framework, this requires an implausibly specific combination of simultaneous conditions. In conclusion, while theoretical expectations cannot be used to confirm observational results, our current understanding of the young population provides a physically compelling argument in favor of the aligned solution for AU Mic c.

\section{Conclusion} \label{sec:summary}

We conducted new RM measurements for AU Mic, a remarkably nearby and young planetary system. We observed both planets, AU Mic b and c, during their transits on two consecutive nights. Our dataset includes Magellan PFS RVs, together with simultaneous CHEOPS and ground-based photometric observations. Our analysis confirms the aligned orbit of AU Mic b, yielding a projected stellar obliquity of $\lambda = 1^\circ \pm 12^\circ$, consistent with previous independent measurements. For AU Mic c, we identify two possible solutions: we slightly favor an aligned orbit with $\lambda_c = -10^\circ \pm 16^\circ$, but cannot rule out a misaligned, nearly polar orbit with $\lambda_c = {87^\circ}_{-29^\circ}^{+36^\circ}$.

While back-to-back observations were intended to compare RM signals under similar activity levels, the results were less than ideal. Rapid stellar rotation caused significant baseline evolution in both photometry and RVs, even within two nights. Moreover, the primary challenge remained the frequent and stochastic transient events, such as flares. At present, neither back-to-back nor repeated observations can efficiently mitigate the impact of such contamination.

To definitively resolve the orbital architecture of the AU Mic system, future observations would ideally include contemporaneous coverage from multiple channels, similar to this work, and need fortuitous windows during quieter stellar conditions. Alternatively, improving our understanding and modeling of stellar activity may prove more effective, benefiting not only future AU Mic observations but also studies of other active stars. Our rich dataset, combining spectroscopy and photometry across two consecutive nights, provides a valuable testbed for developing and evaluating techniques to model stellar jitter.

\begin{acknowledgments}
We thank the anonymous referee for a thorough review and insightful comments, particularly regarding the formation and dynamical perspectives for the Section \ref{subsec:dis_form}.

We thank Tianjun Gan, Huanyu Teng, and Sarah Blunt for their insightful suggestions on this work.

We thank Leon Bewersdorff, Nick Hardy, and Joewie Koh for their help with the ground-based photometric observations. 

Based on observations collected with the PFS/Magellan Clay telescope at Las Campanas Observatory, Chile, under the programme allocated by the Chilean Telescope Allocation Committee (CNTAC), no: 2024B-37.

CHEOPS is an ESA mission in partnership with Switzerland with important contributions to the payload and the ground segment from Austria, Belgium, France, Germany, Hungary, Italy, Portugal, Spain, Sweden, and the United Kingdom. The CHEOPS Consortium would like to gratefully acknowledge the support received by all the agencies, offices, universities, and industries involved. Their flexibility and willingness to explore new approaches were essential to the success of this mission. CHEOPS data analysed in this article are available in the CHEOPS mission archive (\url{https://cheops.unige.ch/archive_browser/}).

This work makes use of observations from the Las Cumbres Observatory global telescope network. This paper is based on observations made with the Las Cumbres Observatory's education network telescopes that were upgraded through generous support from the Gordon and Betty Moore Foundation. This paper is based on observations made with observatory time provided to Boyce Research Initiatives and Education Foundation by the Las Cumbres Observatory through its Global Sky Partners program. 

This work is supported by the China-Chile Joint Research Fund (CCJRF No.2301) and the Chinese Academy of Sciences South America Center for Astronomy (CASSACA) Key Research Project E52H540301. ZGu was supported by the National Research and Development Agency (ANID) through the FONDECYT Iniciación project Grant No.\ 11260176. ZL and SXW acknowledge support from NSFC grant 12273016.

Gy.M.Sz. acknowledges support from SNN-147362 and the ADVANCED-153410 of the National Research, Development and Innovation Office (NKFIH, Hungary), and the ESA PRODEX Experiment Agreements No.\ 4000137122 No.\ 4000149203. 

ZGa acknowledges support from the ESA PRODEX projects Nos. 4000137122, 4000149202, and 4000149203 between ELTE University and the European Space Agency, as well as from the VEGA grant of the Slovak Academy of Sciences No. 2/0033/26, the Slovak Research and Development Agency contract No. APVV-24-0160, the support from SNN-147362 and the ADVANCED-153410 of the National Research, Development and Innovation Office (NKFIH, Hungary), and the support of the city of Szombathely. This work was supported by the bilateral mobility project No. NKM2024-37/HAS-SAS-2024-3. 

ABr acknowledges support by the SNSA.

LW acknowledges the grant (12573089) of the National Natural Science Foundation of China.

GSc and GBr acknowledge support from CHEOPS ASI-INAF agreement n. 2019-29-HH.0.

O.D.S.D. acknowledges support from e-CHEOPS (PEA No 4000142255) and from Fundação para a Ciência e a Tecnologia (FCT) and Fundo Europeu de Desenvolvimento Regional (FEDER) via COMPETE2020 (research grants: UIDB/04434/2020, UIDP/04434/2020, 2022.06962.PTDC).

ML acknowledges support of the Swiss National Science Foundation under grant number PCEFP2\_194576. The contribution of ML has been carried out within the framework of the NCCR PlanetS supported by the Swiss National Science Foundation under grant 51NF40\_205606.

PM acknowledges support from STFC research grants ST/R000638/1, ST/Y002563/1and UKRI1193 and UK Space agency grant UKRI966.

The work of HPO has been carried out within the framework of the NCCR PlanetS supported by the Swiss National Science Foundation under grants 51NF40\_182901 and 51NF40\_205606.

\end{acknowledgments}





%
\facilities{Magellan: Clay II: PFS, CHEOPS, LCOGT: 0.35m, VLT: ESPRESSO, Einstein Probe}

\software{PIPE \citep{Brandeker2024}, AstroImageJ \citep{Collins2017}, BANZAI \citep{McCully2018}, TAPIR \citep{Jensen2013}, starry \citep{Luger2019_starry}, batman \citep{Kreidberg2015_batman}, dynesty \citep{Speagle2020_dynesty1, Koposov2024_dynesty2}, rmfit \citep{Stefansson2022_rmfit}, celerite \citep{Foreman-Mackey2017_celerite}, numpy \citep{Harris2020_numpy}, pandas \citep{mckinney-proc-scipy-2010_pandas2,reback2020_pandas}, scipy \citep{2020SciPy-NMeth_scipy}, astropy \citep{Astropy2013_astropy1,Astropy2018_astropy2,Astropy2022_astropy3}.
          }


\appendix

\section{Spot modeling} \label{sec_app:spot}

To model the large-alplitude long-term variations in light curves, we implement a simple spot model using \texttt{starry} \citep{Luger2019_starry} incorporating spherical harmonics up to the 3rd order. We assume a stellar inclination of $90^\circ$, a rotation period of 4.86 days \citep{Donati2025}, and quadratic limb-darkening coefficients derived for TESS band \citep{Wittrock2023_AU_Mic_TTV}. We fit this model to the long-term ground-based photometry and the CHEOPS light curves, after masking transits and flares. We use the linear solver implemented in \texttt{starry} to solve for the spherical harmonic coefficients and minimize the likelihood to find the best-fit offsets and jitter for each light curve. 

\begin{figure}
\centering
\includegraphics[width=0.49\textwidth]{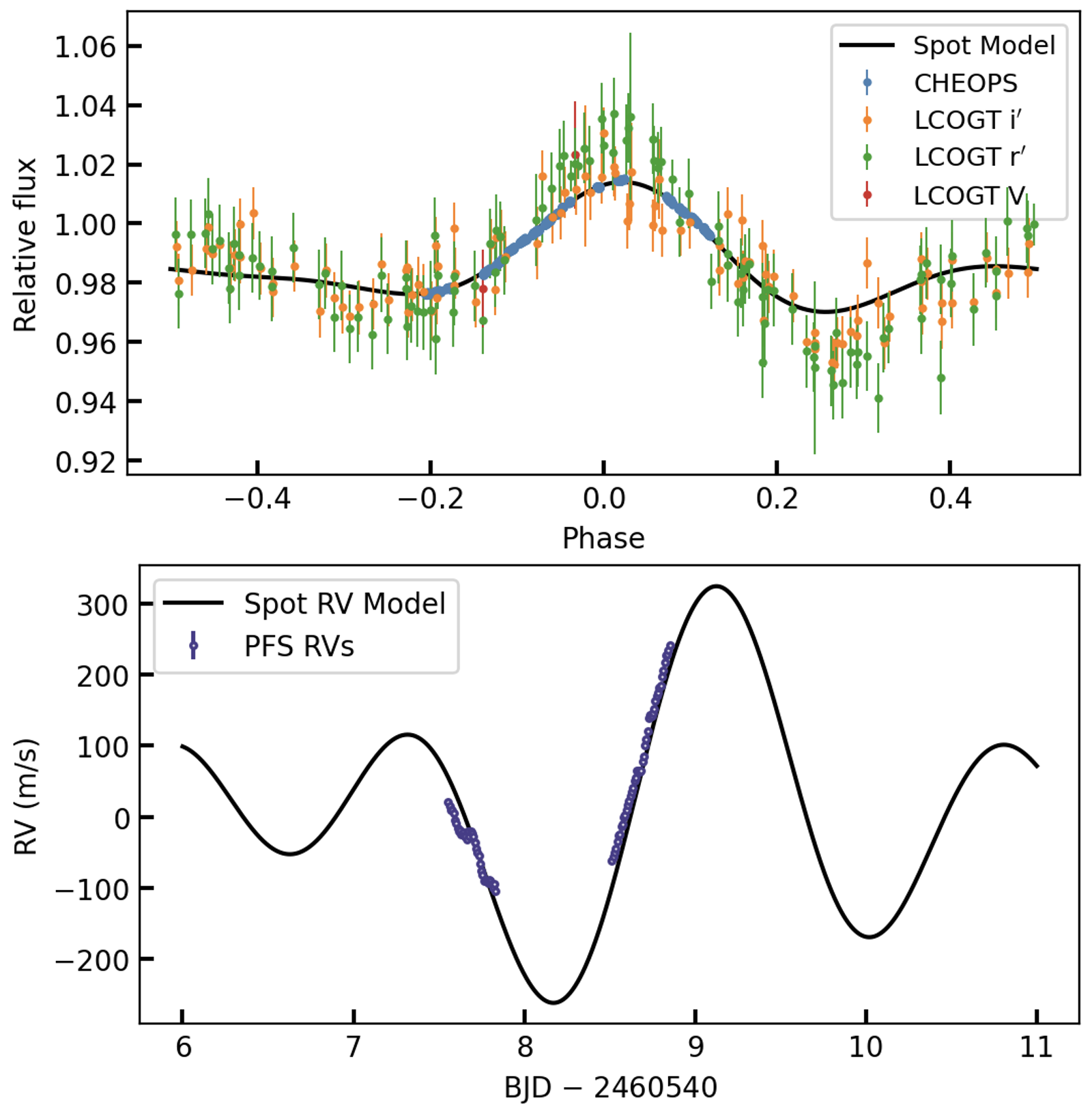}
\caption{\textit{Upper panel:} Long-term ground-based photometric observations and the CHEOPS light curve, plotted with the best-fit \texttt{starry} spot model. The data are phase-folded to the rotation period of 4.86 days. We employ spherical harmonics up to the 3rd degree to construct the spot model. \textit{Lower panel:} The predicted RV variation derived from the best-fit spot model compared to the observed PFS RVs. The spot model can qualitatively match the variation trend of the observed data.} 
\label{fig:spot_model}
\end{figure}

The best-fit model explains the photometric variations well. Notably, the predicted RV variations from this model also qualitatively match the observed RV variations (Appendix Fig.~\ref{fig:spot_model}).

However, for the purpose of modeling the baseline for subsequent transit fitting, this spot model does not perform better than a low-order polynomial (according to $\log \mathcal{Z}$ comparisons). Model settings, such as orders of spherical harmonics or the spot number and shape, cannot be uniquely or robustly constrained by our data or current theoretical frameworks. Fine-tuning these spot parameters could provide a better fit to the data beyond the $\sim$1~ppt level, but this would introduce degeneracies in fitting the transit signal, as 1~ppt is the precision required for transit analysis. Thus, the spot model is not physically reliable under such precision but introduces significantly more free parameters than polynomials. Therefore, our simplified spot model serves primarily as a qualitative demonstration that long-term variability can be approximated by a smooth polynomial fit over a relatively short timescale ($\sim$0.5 days). A similar argument applies to the RV data as well. Consequently, we adopt low-order polynomial models for transit and RM modeling.

\section{Chromatic RV analysis} \label{sec_app:chromatic_rv}

Analyzing RVs across different spectral lines or wavelength ranges can help identify the origins of stellar activities and potentially mitigate their contamination. For example, \cite{Yu2025_AU_Mic_c_RM} conducted a dedicated line-by-line RV analysis on ESPRESSO data to suppress the RV variations by flares. Similar line-by-line analyses are not feasible for PFS, as it employs an iodine cell for wavelength calibration. While PFS covers wavelengths from 391 to 734 nm, RVs were derived only from the iodine absorption region ($\sim$500–600 nm). However, the PFS pipeline first measures the RVs in predefined $\sim$2–3 \AA\ wide spectral chunks and then combines them to yield a single RV measurement. These chunk-by-chunk RVs still allow for a degree of chromatic RV analysis.

Similar to the approach adopted by \cite{Yu2025_AU_Mic_c_RM}, we examine the RVs for each chunk by eye and identify 10 chunks where the RVs show variations with a similar shape but a much larger amplitude (Fig.~\ref{fig:rv_chunk}). When cross-matching with the activity-sensitive line list from \cite{Wise2018}, we find that the top 8 of these 10 chunks contain activity-sensitive lines, most of which are Mg I and Fe I lines.

We try to mitigate the variations by excluding these specific chunks. However, the resulting combined RVs maintained the same anomalous behavior. This suggests that the stellar activity affects other spectral lines as well, albeit with lower amplitudes. We further attempted to use the activity-sensitive line list from \cite{Yu2025_AU_Mic_c_RM} to estimate the sensitivity of each PFS chunk. After excluding the most sensitive chunks, we were still unable to cleanly remove these anomalous variations.

\section{Short Timescale Variability in AU Mic} \label{sec_app:variability}

\begin{figure}
\centering
\includegraphics[width=0.49\textwidth]{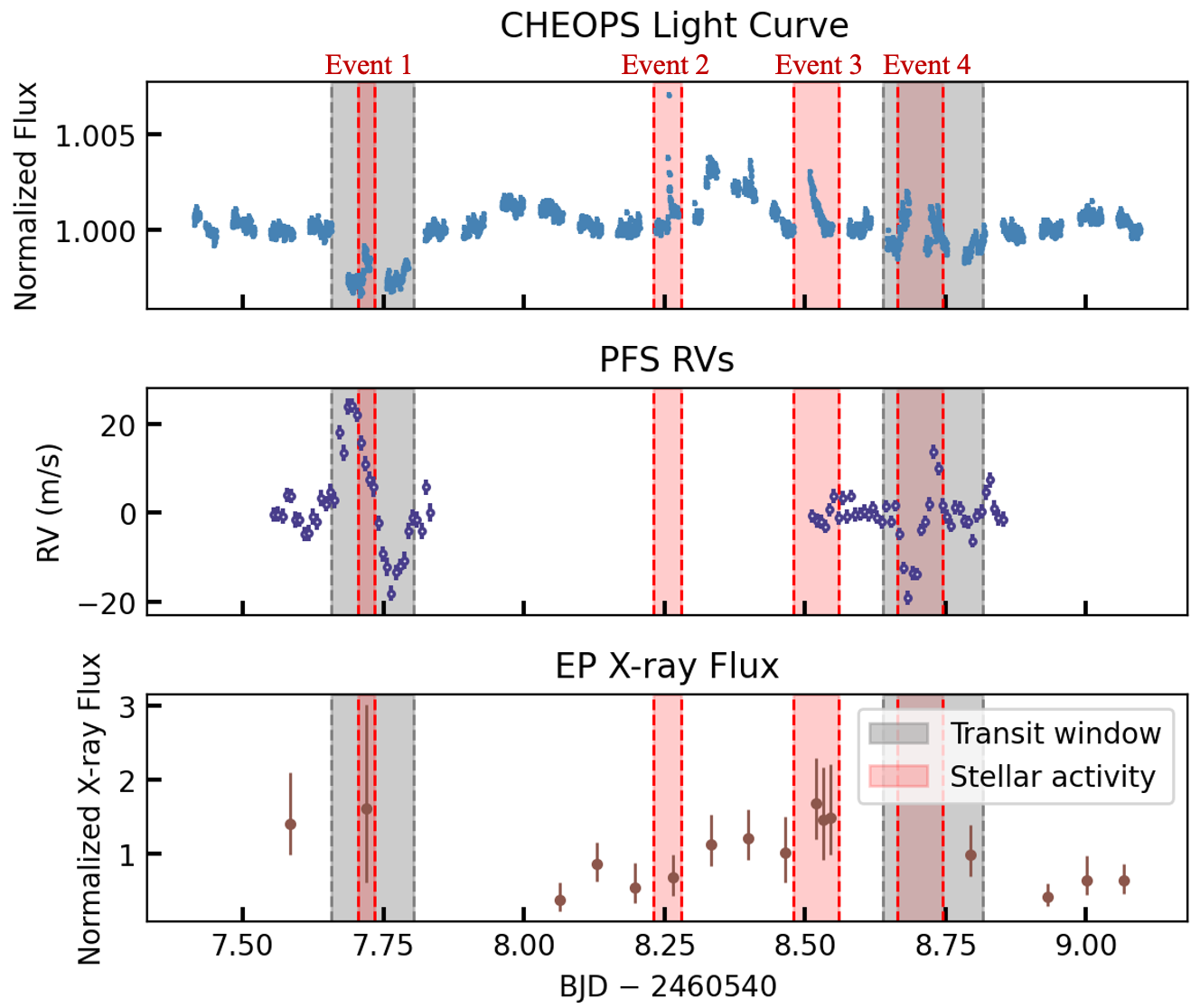}
\caption{CHEOPS light curves, PFS RVs, and EP X-ray time series observed between Aug 24 and 26, 2024. For clarity, long-term stellar rotation trends have been removed from the PFS RVs and CHEOPS light curves. Gray shaded regions denote the transit window for AU Mic b and c. Red shaded regions highlight four stellar activity events, which are further discussed in Appendix \ref{sec_app:variability}.} 
\label{fig:app_LC+RV+EP}
\end{figure}

Short-timescale stellar variability (timescales of order an hour and below) is the dominant source of uncertainty in both our photometry and radial velocities (RVs). Below we summarize the main variability episodes in our time series (Fig.~\ref{fig:app_LC+RV+EP}) and outline plausible physical interpretations.

\paragraph{Event 1: in-transit brightening during AU~Mic~b (JD~2460547.73).}
The photometric morphology and amplitude are consistent with a spot-crossing anomaly. The duration implies an occulted spot with an area of order $\sim$0.3\% of the visible stellar hemisphere, comparable to spot sizes inferred previously from spot-crossing events for AU~Mic \citep[e.g.,][]{Plavchan2020}. No RV perturbation is detected during the anomaly --- this is expected if the occultation occurs near mid-transit, i.e., near the stellar disk center, where the local rotational velocity projected along the line of sight is small.
Alternatively, the event could be a low-energy (micro)flare. We detect a slight contemporaneous enhancement in H$\alpha$ (Figure~\ref{fig:raw_data}), but its amplitude is comparable to pre-event fluctuations, so the evidence for a flare interpretation is weak.

\paragraph{Events 2--3: flaring between the AU~Mic~b and c transits (JD~2460548.25 and 2460548.50).}
Both events show classic white-light flare morphologies (rapid rise followed by slower decay), with additional smaller brightenings in between that could represent unresolved microflares \citep[e.g., solar microflare-like behavior;][]{Battaglia2021}. RV coverage is only available for part of Event 3, where the available RVs sample the decay phase (the first 4--5 RV points on the second night) and show no obvious variability. In contrast, activity tracers (both $S_{\mathrm{HK}}$ and H$\alpha$) exhibit clear decay signatures correlated with the photometric flare decay, supporting the flare interpretation. X-ray monitoring by the \emph{Einstein Probe} \citep{Yuan2022_EP1,Yuan2025_EP2} in the 0.5--4~keV band also shows elevated emission between Events~2 and 3 (a factor of $\gtrsim$2 above quiescence), further indicating enhanced magnetic activity during this interval and corroborating with the flare (and microflare) interpretation.

\paragraph{Event 4: activity complex during the AU~Mic~c transit (JD~2460548.7).}
During the transit of AU~Mic~c we observe a $\sim$2~hr sequence of brightenings. Unlike Event~1, the flux rises above the out-of-transit baseline, ruling out a pure spot-crossing origin. The photometric structure is instead consistent with a cluster of microflares, and the contemporaneous variations in $S_{\mathrm{HK}}$ and (especially) H$\alpha$ are pronounced. This episode produces the largest RV disturbance in our data, with a blueshift followed by a redshift and a peak-to-peak amplitude of $\sim$20\,m\,s$^{-1}$ (stronger in magnetically sensitive lines; Figure~\ref{fig:rv_chunk}). Similar activity-indicator and RV behavior has been reported during flares observed in AU~Mic transit spectroscopy \citep[e.g.,][]{Palle2020_AU_Mic_b_RM,Yu2025_AU_Mic_c_RM}. 

\paragraph{Plausible explanations for Event 4} One plausible explanation of the ``blueshift-then-redshift'' signatures in this work and in ESPRESSO data is the transient appearance and disappearance of redshifted chromospheric emission components superposed on photospheric absorption lines, which would naturally be more prominent in chromospheric diagnostics. Analogous behavior has been reported in the H$\alpha$ core of V830~Tau \citep{Donati2017} and is broadly consistent with recent flare-driven line-formation models \citep{Monson2024}. Alternatively, these signatures could indicate coronal mass ejections (CMEs). Specifically, the “blueshift-then-redshift” pattern could arise if a flare-associated prominence eruption introduces a transient blueshifted component, followed by a redshifted component as part of the ejected material drains back, with both components perturbing the inferred RV through activity-driven line-profile distortions.  However, we do not observe correspondingly strong RV variations in the chunk covering the Fe~5303~\AA\ line, a characteristic line for solar CMEs \citep{Priyal2025}.

The contrast between Event~4 (strong RV response) and Events~1 and 3 (no measurable RV response) underscores that photometric flare strength alone does not uniquely predict the impacts on RVs. Geometric effects (e.g., flares near the limb producing a smaller line-of-sight velocity component) may suppress RV signatures, and some flares may deposit insufficient energy in the lower atmosphere to measurably distort the line profiles \citep{Hudson2006,Fletcher2011}.

The apparent conflicting RV signatures of different flares are not surprising, as their short timescale requires high-cadence, high-precision RV data to capture, which have been sparse so far for active M dwarfs. While larger flares are easy to identify and could have a large effect on RVs \citep[$\sim$100 m/s;][]{Reiners2009}, in general, the results have been inconclusive for small and medium flares \citep[e.g.,][]{Barnes2014,Anglada-Escude2016_Proxima,Pavlenko2017}. \cite{Anglada-Escude2016_Proxima} captured a relatively big flare in their high-cadence HARPS program on Proxima Centauri, but contrary to \cite{Reiners2009}, the big flare caught in action in Proxima did not seem to have any effect beyond a couple m/s. On the other hand, flares from young active stars like V830 Tau have shown RV effects as large as 0.3 km/s \citep{Donati2017}.

We caution that a complete, predictive theory for broadband spectral-line and RV perturbations during flares (on the Sun and other stars) remains lacking. Progress will require simultaneous high-cadence, high-precision RV spectroscopy and continuous photometric monitoring to map flare properties to their spectroscopic imprints.

\section{RM fitting results for other models} \label{sec_app:RM}

\begin{figure*}
\centering
\includegraphics[width=0.95\textwidth]{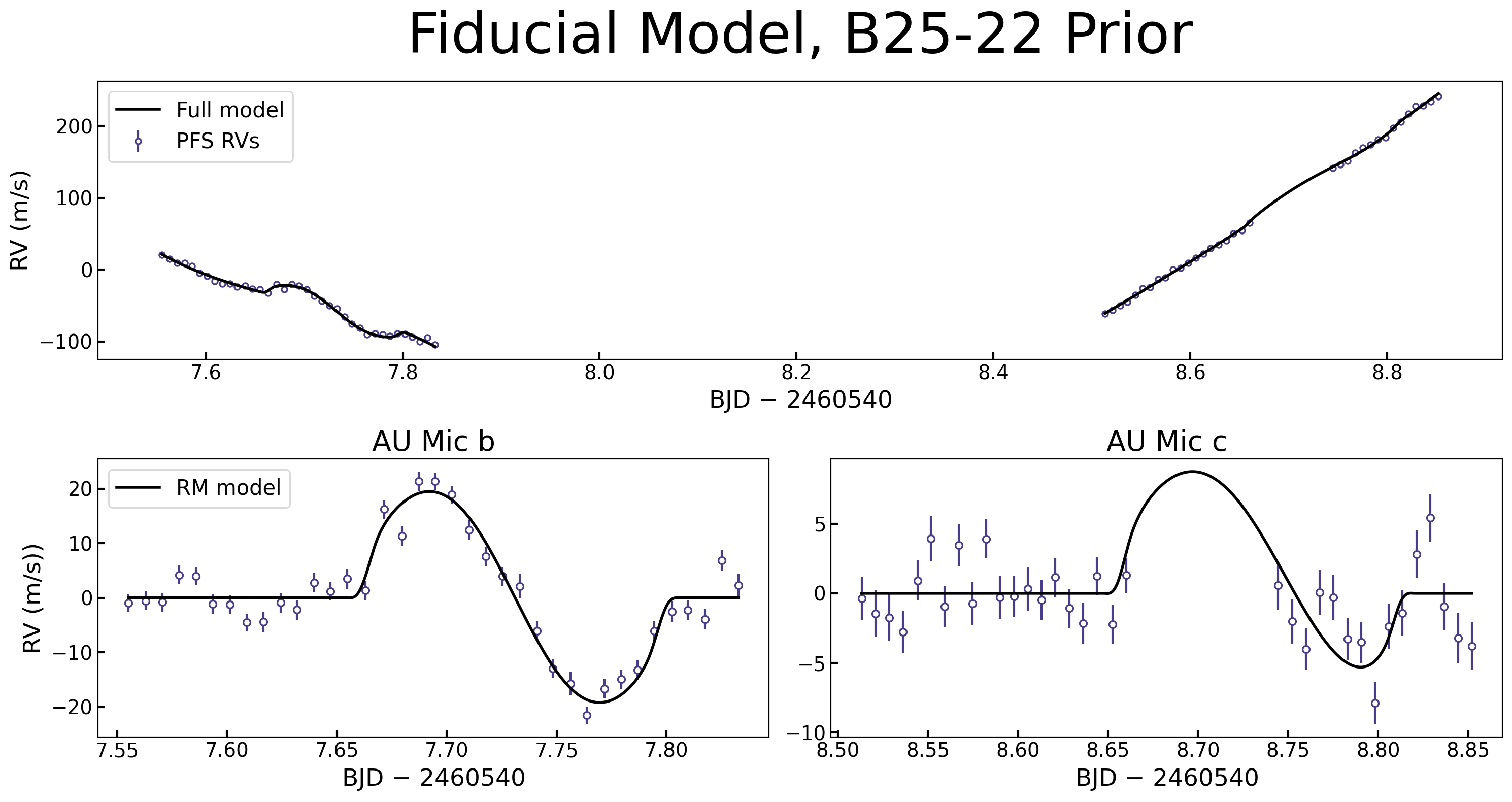}
\caption{Figure set for the best-fit modeling results of the RM effect for AU Mic b and c with PFS RVs. Each figure presents one model tested in Section \ref{sec:RM}. The complete figure set (17 images) is available in the online journal.}
\label{fig_set:RM}
\end{figure*}

This appendix displays the figure set (Fig.~\ref{fig_set:RM}) of the best-fit modeling results of the RM effect.

\section{Priors and posteriors for transit and RM modeling} \label{sec_app:corner}

\begin{figure*}
\centering
\includegraphics[width=0.99\textwidth]{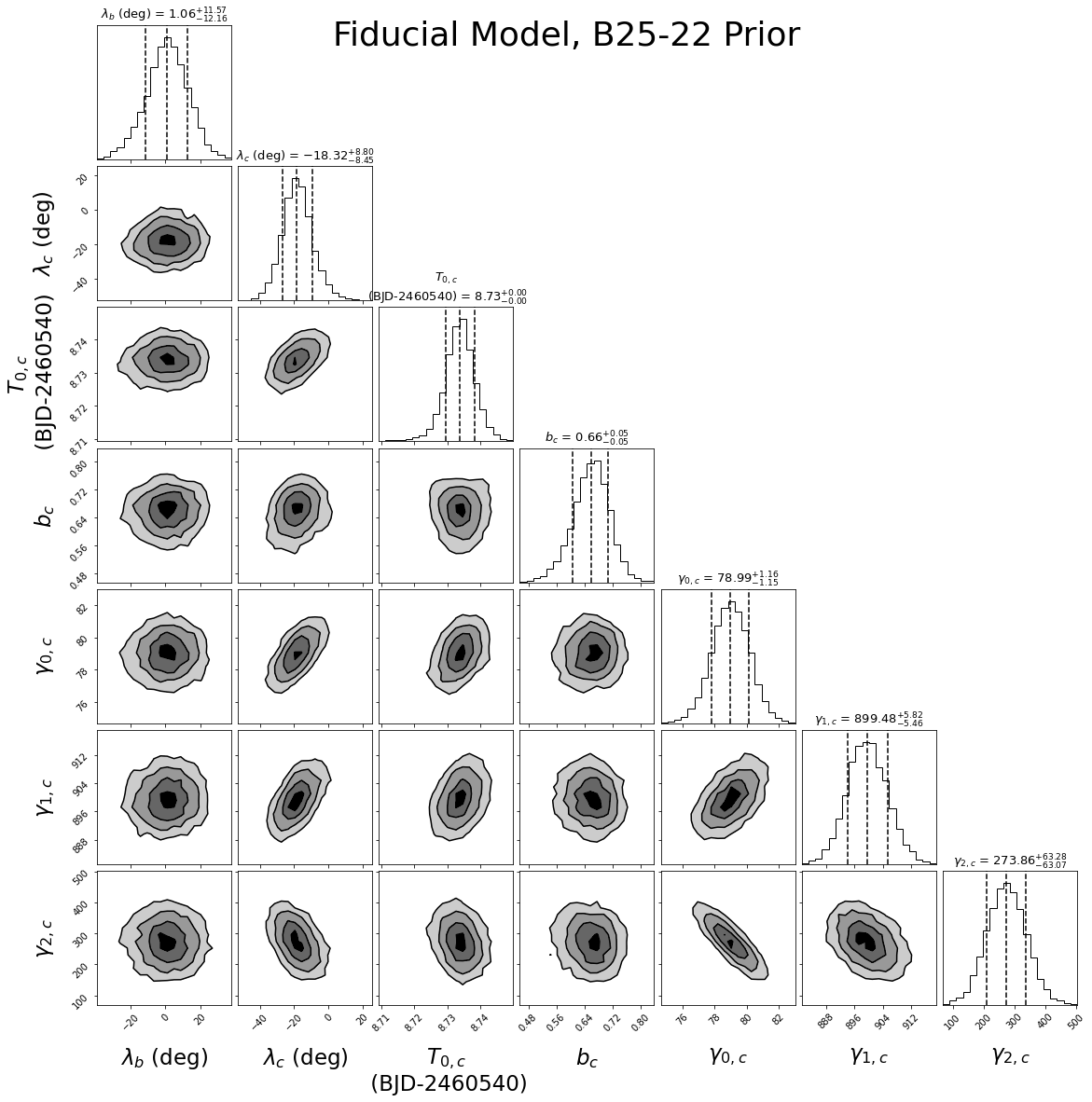}
\caption{The figure set for the posterior distributions of the RM model fitting. For clarity, we only show $\rm \lambda_b$, $\rm \lambda_c$, and important parameters related to $\rm \lambda_c$. Each figure presents one model tested in Section \ref{sec:RM}. The complete figure set (9 images) is available in the online journal.}
\label{fig_set:corner}
\end{figure*}

This appendix concludes the prior settings for transit and RM modeling in Table \ref{tab:transit} and \ref{tab:RM}. The figure set for the posterior distributions of the RM fittings is shown in Fig.~\ref{fig_set:corner}. Complete posterior tables for the transit and RM modeling are available on Zenodo: \dataset[doi:10.5281/zenodo.18910646]{https://doi.org/10.5281/zenodo.18910646}

\begin{table*}
    \centering
    \caption{Prior settings for the transit fit of AU Mic b and c. $\mathcal{N}$($\mu$, $\sigma$) means a normal prior with mean $\mu$ and standard deviation $\sigma$. $\mathcal{U}$(a, b) stands for a uniform prior ranging from a to b. $\mathcal{J}$(a, b) stands for a log-uniform prior ranging from a to b. $\mathcal{TN}$($\mu$, $\sigma$, a, b) stands for a truncated normal prior with mean $\mu$, standard deviation $\sigma$, lower bound $a$, and upper bound b. The W23 prior is based on \cite{Wittrock2023_AU_Mic_TTV}, and the B25-22 and B25-23 priors are based on 2022 and 2023 values of \cite{Boldog2025_CHEOPS_TTV}, respectively.}
    \begin{tabular}{lccc}
    \hline\hline
   Parameter &  \multicolumn{3}{c}{Prior} \\
    & W23 & B25-22 & B25-23 \\
    \hline
    AU Mic b \\
    $P$ (days) & Fixed, 8.46308 & Fixed, 8.46314 & Fixed, 8.46314 \\
    $T_0$ (BJD$-2460540$) & $\mathcal{U}$(7.7, 7.8) & $\mathcal{U}$(7.7, 7.8) & $\mathcal{U}$(7.7, 7.8) \\
    $R_p/R_*$ & $\mathcal{N}$(0.0488, 0.001) & $\mathcal{N}$(0.047, 0.0008) & $\mathcal{N}$(0.0517, 0.0011)\\
    $a/R_*$ & $\mathcal{N}$(18.8, 0.6) & $\mathcal{N}$(18.8, 0.5) & $\mathcal{N}$(18.2, 0.5)\\
    $b$ & $\mathcal{TN}$(0.13, 0.1, 0, 1) & $\mathcal{TN}$(0.29,0.1,0,1) & $\mathcal{TN}$(0.39,0.1,0,1)\\
    $e$ & $\mathcal{TN}$(0.1, 0.1, 0, 0.5) &  Fixed, 0 & Fixed, 0 \\
    $\omega$ (deg) & $\mathcal{U}$(-180, 180) &  Fixed, 90 & Fixed, 90 \\
    $\gamma_{\rm 0,CHEOPS}$$^{[2]}$ & \multicolumn{3}{c}{$\mathcal{U}$(0.9, 1.1)}  \\
    $\gamma_{\rm 1,CHEOPS}$ & \multicolumn{3}{c}{$\mathcal{U}$(-0.1, 0.1)} \\
    $\gamma_{\rm 2,CHEOPS}$ & \multicolumn{3}{c}{$\mathcal{U}$(-1, 1)} \\
    $t_{\rm base,CHEOPS}$ (BJD$-2460540$) & \multicolumn{3}{c}{Fixed, mean of fitted CHEOPS data}  \\
    $T_{\rm 14}$$^{[3]}$ (days) & $\mathcal{N}$(0.1455, 0.001) & $\mathcal{N}$(0.144,0.0009) & $\mathcal{N}$(0.145,0.0025) \\
    \hline
    AU Mic c \\
    $P$ (days) & Fixed, 18.85969 & Fixed, 18.858819 & Fixed, 18.858827  \\
    $T_0$ (BJD$-2460540$) & $\mathcal{U}$(8.68, 8.85) & $\mathcal{U}$(8.68, 8.85) & $\mathcal{U}$(8.68, 8.85)  \\
    $R_p/R_*$ & $\mathcal{N}$(0.0311, 0.0028) & $\mathcal{N}$(0.0354, 0.0016) & $\mathcal{N}$(0.0309, 0.0033)  \\
    $a/R_*$ & $\mathcal{N}$(32.05, 1.0) & $\mathcal{N}$(30.6, 1.0) & $\mathcal{N}$(30.7, 1.0)  \\
    $b$ & $\mathcal{TN}$(0.3, 0.2, 0, 1) & $\mathcal{TN}$(0.61, 0.12, 0, 1) & $\mathcal{TN}$(0.66, 0.2, 0, 1)  \\
    $e$ & $\mathcal{TN}$(0.1, 0.1, 0, 0.5) & Fixed, 0 & Fixed, 0 \\
    $\omega$ (deg) & $\mathcal{U}$(-180, 180) &  Fixed, 90 & Fixed, 90  \\
    $\gamma_{\rm 0,CHEOPS}$ & \multicolumn{3}{c}{$\mathcal{U}$(0.9, 1.1)}  \\
    $\gamma_{\rm 1,CHEOPS}$ & \multicolumn{3}{c}{$\mathcal{U}$(-0.1, 0.1)}  \\
    $\gamma_{\rm 2,CHEOPS}$ & \multicolumn{3}{c}{$\mathcal{U}$(-1, 1)}  \\
    $\gamma_{\rm 3,CHEOPS}$ & \multicolumn{3}{c}{$\mathcal{U}$(-1, 1)}  \\
    $\gamma_{\rm 4,CHEOPS}$ & \multicolumn{3}{c}{$\mathcal{U}$(-1, 1)}  \\
    $t_{\rm base,CHEOPS}$ (BJD$-2460540$) & \multicolumn{3}{c}{Fixed, mean of fitted CHEOPS data} \\
    $\gamma_{\rm 0,LCOGT1}$ & \multicolumn{3}{c}{$\mathcal{U}$(0.9, 1.1)}  \\
    $\gamma_{\rm 1,LCOGT1}$ & \multicolumn{3}{c}{$\mathcal{U}$(-0.1, 0.1)}  \\
    $\gamma_{\rm 2,LCOGT1}$ & \multicolumn{3}{c}{$\mathcal{U}$(-1, 1)}  \\
    $t_{\rm base,LCOGT1}$ (BJD$-2460540$) & \multicolumn{3}{c}{Fixed, mean of fitted LCOGT data}  \\
    $\gamma_{\rm 0,LCOGT2}$ & \multicolumn{3}{c}{$\mathcal{U}$(0.9, 1.1)}  \\
    $\gamma_{\rm 1,LCOGT2}$ & \multicolumn{3}{c}{$\mathcal{U}$(-0.1, 0.1)}  \\
    $\gamma_{\rm 2,LCOGT2}$ & \multicolumn{3}{c}{$\mathcal{U}$(-1, 1)}  \\
    $t_{\rm base,LCOGT2}$ (BJD$-2460540$) & \multicolumn{3}{c}{Fixed, mean of fitted LCOGT data}  \\
    $\sigma_{\rm jit,LCOGT1}$ (ppt) & \multicolumn{3}{c}{$\mathcal{J}$($10^{-3}$, $10$)}  \\
    $\sigma_{\rm jit,LCOGT2}$ (ppt) & \multicolumn{3}{c}{$\mathcal{J}$($10^{-3}$, $10$)}  \\
    $u_{\rm 1,LCOGT}$$^{[4]}$ & Fixed, 0.467 & Fixed, 0.39 & Fixed, 0.60 \\
    $u_{\rm 2,LCOGT}$ & Fixed, 0.186 & Fixed, 0.30 & Fixed, 0.19 \\ 
    $T_{\rm 14}$$^{[2]}$ (days) & $\mathcal{N}$(0.1765,0.01) & $\mathcal{N}$(0.1625,0.013) & $\mathcal{N}$(0.1554,0.027)  \\
    \hline
    Same for AU Mic b and c \\
    $\sigma_{\rm jit,CHEOPS}$ (ppt) & \multicolumn{3}{c}{$\mathcal{J}$($10^{-3}$, $10$)}  \\
    $u_{\rm 1,CHEOPS}$ & Fixed, 0.467 & Fixed, 0.39 & Fixed, 0.60 \\
    $u_{\rm 2,CHEOPS}$ & Fixed, 0.186 & Fixed, 0.30 & Fixed, 0.19 \\    
    \hline\hline 
    \end{tabular}
    \begin{tablenotes}
    \item[1]  [1] The polynomial trend for each instrument is $\sum_n \gamma_{\rm n,ins}(t-t_{\rm base,ins})^n$
    \item[2]  [2] $T_{\rm 14}$ is not a free parameter in the fitting. As our light curves may not cover the ingress or the egress, we add this prior into the likelihood calculation.
    \item[3]  [3] The limb-darkening parameters for both LCOGT observations are set to the TESS values from \cite{Wittrock2023_AU_Mic_TTV}, because the $i'$ band is close to the TESS band, and the TESS data provide the most precise constraints.
    \end{tablenotes}
    \label{tab:transit}
\end{table*}

\begin{table*}
    \centering
    \caption{Prior settings for the RM effect fit of AU Mic b and c. The prior notations are the same as Table \ref{tab:transit}.}
    \begin{tabular}{lcccc}
    \hline\hline
   Parameter &  \multicolumn{3}{c}{Prior} \\
    & W23 & B25-22 & B25-23 \\
    \hline
    AU Mic b \\
    $\lambda$ (deg) & \multicolumn{3}{c}{$\mathcal{U}$(-180, 180)}  \\
    $P$ (days) & Fixed, 8.46308 & Fixed, 8.46314 & Fixed, 8.46314 \\
    $T_0$ (BJD$-2460540$) & \multicolumn{3}{c}{$\mathcal{N}$(7.731, 0.002)} \\
    $R_p/R_*$ & $\mathcal{N}$(0.0488, 0.001) & $\mathcal{N}$(0.047, 0.0008) & $\mathcal{N}$(0.0517, 0.0011)\\
    $a/R_*$ & $\mathcal{N}$(18.8, 0.6) & $\mathcal{N}$(18.8, 0.5) & $\mathcal{N}$(18.2, 0.5)\\
    $b$ & $\mathcal{TN}$(0.13, 0.1, 0, 1) & $\mathcal{TN}$(0.29,0.1,0,1) & $\mathcal{TN}$(0.39,0.1,0,1)\\
    $e$ & $\mathcal{TN}$(0.1, 0.1, 0, 0.5) &  Fixed, 0 & Fixed, 0 \\
    $\omega$ (deg) & $\mathcal{U}$(-180, 180) &  Fixed, 90 & Fixed, 90 \\
    $\gamma_{\rm 0,b}$$^{[1]}$ & \multicolumn{3}{c}{$\mathcal{U}$(-500, 500)}  \\
    $\gamma_{\rm 1,b}$ & \multicolumn{3}{c}{$\mathcal{U}$(-10000, 10000)}  \\
    $\gamma_{\rm 2,b}$ & \multicolumn{3}{c}{$\mathcal{U}$(-10000, 10000)}  \\
    $\gamma_{\rm 3,b}$ & \multicolumn{3}{c}{$\mathcal{U}$(-20000, 20000)}  \\
    $t_{\rm base,b}$ (BJD$-2460540$) & \multicolumn{3}{c}{Fixed, 7.6942} \\
    $T_{\rm 14}$$^{[2]}$ (days) & $\mathcal{N}$(0.1455, 0.001) & $\mathcal{N}$(0.144, 0.0009) & $\mathcal{N}$(0.145, 0.0025) \\
    \hline
    AU Mic c \\
    $\lambda$ (deg) & \multicolumn{3}{c}{$\mathcal{U}$(-180, 180)}  \\
    $P$ (days) & Fixed, 18.85969 & Fixed, 18.858819 & Fixed, 18.858827  \\
    $T_0$ (BJD$-2460540$) & \multicolumn{3}{c}{$\mathcal{N}$(8.723, 0.01)}  \\
    $R_p/R_*$ & $\mathcal{N}$(0.0311, 0.0028) & $\mathcal{N}$(0.0354, 0.0016) & $\mathcal{N}$(0.0309, 0.0033)  \\
    $a/R_*$ & $\mathcal{N}$(32.05, 1.0) & $\mathcal{N}$(30.6, 1.0) & $\mathcal{N}$(30.7, 1.0)  \\
    $b$ & $\mathcal{TN}$(0.3, 0.2, 0, 1) & $\mathcal{TN}$(0.61, 0.12, 0, 1) & $\mathcal{TN}$(0.66, 0.2, 0, 1)  \\
    $e$ & $\mathcal{TN}$(0.1, 0.1, 0, 0.5) & Fixed, 0 & Fixed, 0 \\
    $\omega$ (deg) & $\mathcal{U}$(-180, 180) &  Fixed, 90 & Fixed, 90  \\
    $\gamma_{\rm 0,c}$ & \multicolumn{3}{c}{$\mathcal{U}$(-500, 500)}  \\
    $\gamma_{\rm 1,c}$ & \multicolumn{3}{c}{$\mathcal{U}$(-10000, 10000)}  \\
    $\gamma_{\rm 2,c}$$^{[3]}$ & \multicolumn{3}{c}{$\mathcal{U}$(-10000, 10000)}  \\
    $t_{\rm base,c}$ (BJD$-2460540$) & \multicolumn{3}{c}{Fixed, 8.6771}  \\
    $T_{\rm 14}$$^{[2]}$ (days) & $\mathcal{N}$(0.1765, 0.01) & $\mathcal{N}$(0.1625, 0.013) & $\mathcal{N}$(0.1554, 0.027)  \\ 
    \hline
    $v_* \sin i$ (km/s) & \multicolumn{3}{c}{$\mathcal{U}$(5, 10)}  \\
    $u_1$ & $\mathcal{N}$(0.467, 0.1) & $\mathcal{N}$(0.39, 0.1) & $\mathcal{N}$(0.6, 0.1)  \\
    $u_2$ & $\mathcal{N}$(0.186, 0.1) & $\mathcal{N}$(0.30, 0.1) & $\mathcal{N}$(0.19, 0.1)  \\
    $\sigma_{\rm jit,PFS}$ (m/s) & \multicolumn{3}{c}{$\mathcal{J}$(0.01, 100)}  \\
    $\beta$$^{[4]}$ (km/s) & \multicolumn{3}{c}{$\mathcal{N}$(2.36, 0.5)}  \\
    $t_{\rm exp}$ (s) & \multicolumn{3}{c}{Fixed, 600} \\
    \hline
    GP (Mat\'ern 3/2 kernel)$^{[5]}$\\
    $\log \sigma_{\rm GP}$ & \multicolumn{3}{c}{$\mathcal{U}$(-2, 5)}  \\
    $\log \rho_{\rm GP}$$^{[6]}$ & \multicolumn{3}{c}{$\mathcal{N}$(-4, 0.5) (GP Model 1)}  \\
     & \multicolumn{3}{c}{$\mathcal{N}$(-2.8, 0.2) (GP Model 2)} \\
    \hline\hline 
    \end{tabular}
    \begin{tablenotes}
    \item[1]  [1] The notation b and c stand for the first night and the second night, respectively. The polynomial trend for each night is $\sum_n \gamma_{\rm n}(t-t_{\rm base})^n$
    \item[2]  [2] $T_{\rm 14}$ is not a free parameter in the fitting. We keep the same settings as the transit fit.
    \item[3]  [3] Not used in GP models.
    \item[4]  [4] The thermal velocity parameter introduced in \cite{Hirano2011}, which is related to the instrumental profile of the spectrograph. The prior is based on $R = 127\,000$ for PFS. We also tested priors with different standard deviations ($\sigma$) and uniform priors, and found consistent results for the obliquity constraints.
    \item[5]  [5] Not used in fiducial models.
    \item[6]  [6] We tried two different priors on $\log \rho_{\rm GP}$ (time scale of the GP) in the two GP fits. 
    \end{tablenotes}
    \label{tab:RM}
\end{table*}

\bibliography{references}{}
\bibliographystyle{aasjournalv7}



\end{document}